\documentclass[11pt]{article}

\usepackage[final]{acl}

\usepackage{times}
\usepackage{latexsym}
\usepackage{microtype}
\usepackage{enumitem}
\usepackage{graphicx}
\usepackage[ruled,vlined,linesnumbered]{algorithm2e}
\usepackage{xcolor}
\usepackage{colortbl}
\usepackage{multirow}
\usepackage{subcaption}
\usepackage{booktabs}
\usepackage{array}
\usepackage{enumerate}
\usepackage{caption}
\usepackage{pifont}
\usepackage{wrapfig}
\usepackage[utf8]{inputenc} 
\usepackage[T1]{fontenc}    
\usepackage{hyperref}       
\usepackage{url}            
\usepackage{amsfonts}       
\usepackage{nicefrac}       
\usepackage{microtype}      
\usepackage{tikz}
\usepackage[most]{tcolorbox}
\usepackage{wrapfig}
\usepackage{fontawesome5} 

\tcbuselibrary{listings}
\newtcolorbox{promptbox}{
  colback=gray!5, colframe=black!50,
  listing only,
  listing options={basicstyle=\ttfamily\small, breaklines=true},
  keepspaces=true
}

\usepackage{colortbl}
\definecolor{grey}{rgb}{0.89,0.71,0.57}
\definecolor{pink}{rgb}{1,0.94,1}
\definecolor{purple}{rgb}{0.84,0.78,1}
\definecolor{white}{rgb}{1,1,1}
\definecolor{backred}{RGB}{255, 190, 190}
\definecolor{backblue}{RGB}{210, 230, 250}
\definecolor{mygrey}{RGB}{200,200,200}
\definecolor{codegreen}{rgb}{0,0.6,0}
\definecolor{codegray}{rgb}{0.5,0.5,0.5}
\definecolor{codepurple}{rgb}{0.58,0,0.82}
\definecolor{backcolour}{rgb}{0.95,0.95,0.92}
\definecolor{lightyellow}{RGB}{255, 252, 51}
\definecolor{lightgreen}{RGB}{204, 255, 204}
\definecolor{darkred}{RGB}{139,0,0} 

\hypersetup{
    colorlinks=true,
    linkcolor=red,
    citecolor=cyan,
    filecolor=magenta,
    urlcolor=magenta,
    }

\renewcommand{\cite}{\citep}  

\newtcolorbox{boxK}{
    top=2.2pt,
    bottom=2.2pt,
    left=4.5pt,
    right=4.5pt,
    boxrule = 0pt,
    toprule = 0pt, 
    enhanced,
}

\newcommand{\llmname}[1]{{\fontfamily{pcr}\selectfont {#1}}\xspace}
\newcommand{\dataname}[1]{{\fontfamily{cmtt}\selectfont {#1}}\xspace}

\title{DiffuTester: Accelerating Unit Test Generation for Diffusion LLMs via Mining Structural Pattern}

\author{
Lekang Yang,
Yuetong Liu,
Yitong Zhang,
Jia Li\thanks{Corresponding author} \\
College of AI, Tsinghua University, Beijing, China \\
\texttt{ylk22@mails.tsinghua.edu.cn}, \texttt{23371522@buaa.edu.cn},\\
\texttt{zhangyt26@mails.tsinghua.edu.cn}, \texttt{jia\_li@mail.tsinghua.edu.cn}
}

\def\eg{\emph{e.g.,}\xspace}

\def\rawname{DiffuTester\xspace}
\def\scname{\textsc{\rawname}\xspace}
\def\name{\textsc{\rawname}\xspace}

\begin{document}

\maketitle

\begin{abstract}
Diffusion large language models (dLLMs) enable parallel generation and are promising for unit test generation (UTG), where efficient and large-scale automated testing is essential in software development.
Despite this advantage, their application to UTG is still constrained by a clear trade-off between efficiency and test quality, since increasing the number of tokens generated in each step often causes a sharp decline in the quality of test cases.
To overcome this limitation, we present \name, an acceleration framework specifically tailored for dLLMs in UTG. The motivation of \name is that unit tests targeting the same focal method often share structural patterns. \name employs a novel structural pattern based decoding approach, which dynamically identifies structural patterns across unit tests through their abstract syntax trees and additionally decodes the corresponding tokens, thereby achieving acceleration without compromising the quality of the output.
To enable comprehensive evaluation, we extend the original TestEval benchmark to three programming languages. Extensive experiments on three benchmarks with two representative models show that \name delivers significant acceleration while preserving test coverage. Moreover, \name generalizes well across different dLLMs and programming languages, providing a practical and scalable solution for efficient UTG in software development.
Code and data are publicly available at \url{https://github.com/THU-Agent/DiffuTester}.
\end{abstract}

\section{Introduction}

Recently, Diffusion Large Language Models (dLLMs)~\cite{llada, dream} are emerging as a promising paradigm because of their unique generation paradigm, which naturally supports multi-token prediction and enables faster generation. An increasing number of dLLM models~\cite{dreamcoder, diffucoder} for code generation are emerging as a result of the high efficiency of dLLMs.

Unit testing plays a vital role in software development, ensuring that a functionally discrete program unit (\eg{a method}) behaves correctly and meets the intended design expectations~\cite{UT1, gren2017relation}.
Real-world code repositories are often large, requiring developers to write a substantial number of unit tests (UTs) to cover each statement and branch~\cite{robinson2011scaling, li2006large}, which is extremely time-consuming and labor-intensive~\cite{UT2, shang2025large}.
In this context, efficiently generating unit tests becomes critically important, and dLLMs are particularly well-suited to this requirement.

Although dLLMs are theoretically capable of generating multiple tokens per denoising step, existing studies~\cite{zeng2025treediff, time-consuming1} typically retain only one or two tokens per step to maintain generation quality.
Our preliminary experiments show that naively increasing the number of retained tokens substantially accelerates decoding but sharply degrades the quality of generated unit tests (see Appendix~\ref{sec:steps}).
This speed-quality trade-off motivates the need for a UTG-specific acceleration method that can decode more tokens per step while preserving test quality.

\begin{figure*}[!t]
    \centering
    \includegraphics[width=0.8\textwidth]{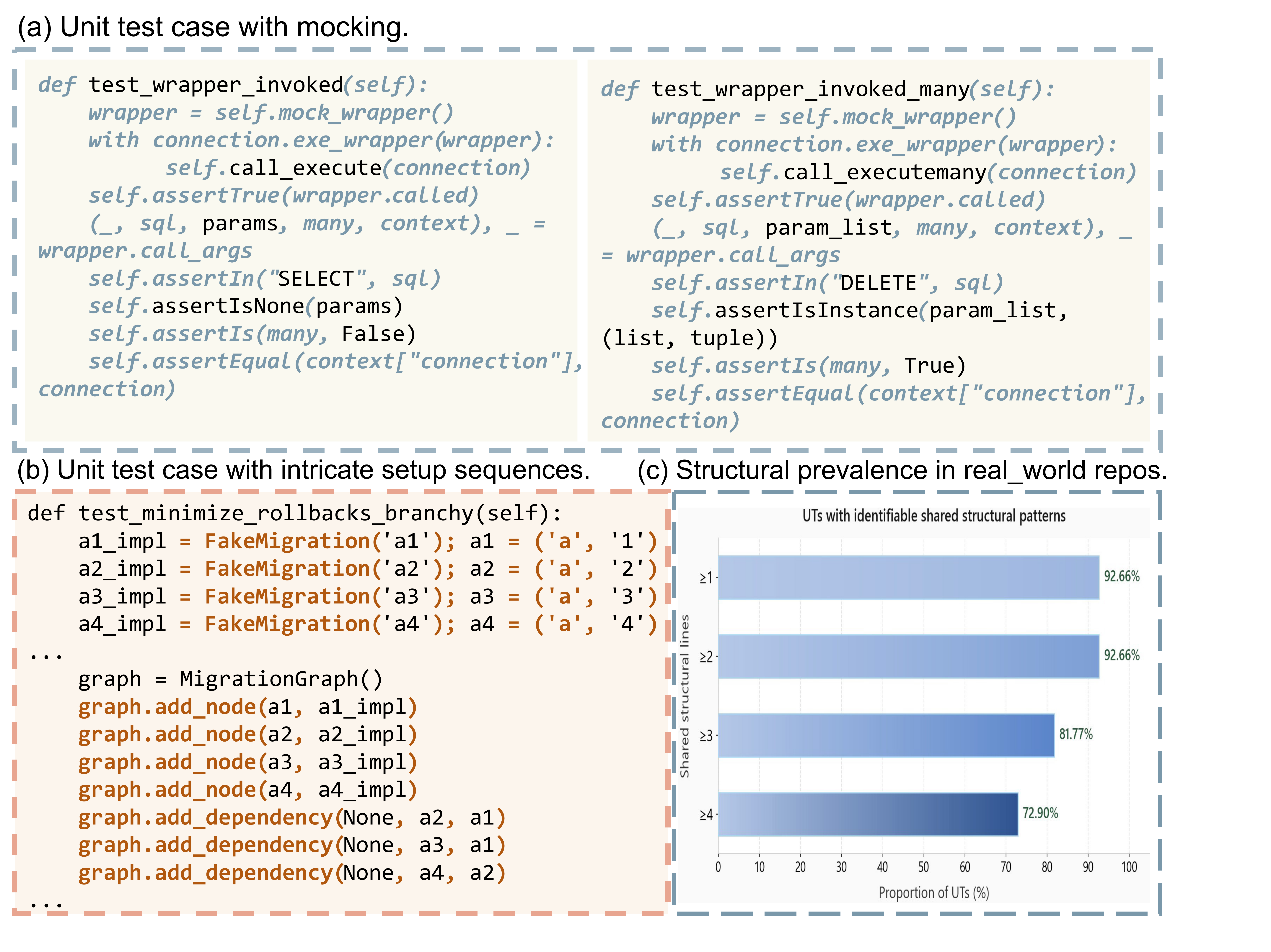}
    \caption{Real-world unit tests exhibit recurring structural patterns that can be exploited by \name.}
    \label{fig:similarity}
\end{figure*}

To address this limitation, we propose \name to accelerate UTG for dLLMs.
Our method is inspired by an observation that test cases targeting the same focal method often share similar structural patterns.
In Figure~\ref{fig:similarity}(a), two mocking-based test cases exhibit nearly identical syntactic structures, while differing only in values and constants.
Figure~\ref{fig:similarity}(b) further shows that even within a single complex test case with intricate setup sequences, repeated structural patterns frequently appear in setup and execution logic.
To further validate this motivation, we randomly sample 100 test files from real-world repositories, containing 4,646 unit tests in total.
Using the AST-merge algorithm in \name, we find that 92.66\% of UTs contain at least two shared structural lines (as illustrated in Figure~\ref{fig:similarity}(c)).
These results indicate that structural patterns are highly prevalent in real-world unit tests and can be effectively captured by \name, thereby supporting our motivation.

To exploit structural patterns, \name dynamically mines such patterns at each intermediate step of the diffusion process and additionally decodes the tokens corresponding to these patterns. Therefore it reduces the number of denoising steps by retaining more reliable tokens at each step, which leads to the core speedup of \name.
To enable this acceleration, \name must first solve the problem of identifying structurally reliable tokens from incomplete test outputs.
Abstract syntax trees (ASTs) provide a natural representation for separating syntactic structure from instance-specific lexical content in code.
Based on this property, \name parses the generated code into multiple ASTs and extracts the common nodes across them, which we regard as indicative of structural patterns inherent to the test cases.

We perform extensive experiments to validate the effectiveness of our proposed \name on the \dataname{TestEval} benchmark~\cite{testeval} (whose focal methods are all implemented in Python) and new benchmarks collected by us (\dataname{TestEval-C++} and \dataname{TestEval-Java}) with two representative dLLMs, including \llmname{Dream}~\cite{dream} and \llmname{DiffuCoder}~\cite{diffucoder}.
The results show that:
\ding{182} \name accelerates UTG by up to $2\times$--$3\times$ and effectively reduces computational cost.
\ding{183} \name preserves the maximum achievable test coverage while improving decoding efficiency.
\ding{184} \name is training-free and generalizes across different dLLMs, programming languages, code generation tasks, and real-world repository-level UTG scenarios, achieving about $1.9\times$ speedup on \dataname{Tests4Py}.

\section{Related Work}

Diffusion language models have recently become a focal point in AI research~\cite{sahoo2024simple, discreteVLA, zhang2026lookaheadthenverifyreliableconstraineddecoding, li2026diffuguardintrinsicsafetylost}, especially discrete diffusion language models~\cite{llada, gemini_diffusion, dream, sdar}. More and more dLLMs are also applied to code generation as a result of their high generation efficiency~\cite{diffucoder, dreamcoder}.

Various training-free strategies have been developed to accelerate the dLLMs generation process. One major class of acceleration methods is based on KV-cache mechanisms, such as DLLM-Cache~\cite{dLLM-Cache} and Sparse-dLLM~\cite{song2025sparse}. Another class of acceleration methods is focused on advanced sampling techniques~\cite{wei2025accelerating, ben2025accelerated, israel2025accelerating,yang2026improvingsamplingmaskeddiffusion}.
However, these general-purpose acceleration techniques are not tailored to UTG and may result in degradation of generation quality, whereas \name leverages UTG-specific structural regularities to improve decoding efficiency while maintaining test quality.

\section{\rawname}
\label{sec:methodology}

We propose \name, an approach designed to accelerate dLLMs
in unit test generation.
The key idea is to mine shared structural patterns observed in multiple unit test cases generated at intermediate inference steps and then leverage them to guide token generation.
We first provide a brief overview of the overall acceleration procedure (Section~\ref{sec:overview}). Then we provide an in-depth explanation of how structural patterns are mined, which forms the core of our approach (Section~\ref{sec:core}). Finally, we introduce several additional techniques to further accelerate inference and ensure the quality of the generated test cases (Section~\ref{sec:additional}). An overview of \name is presented in Figure~\ref{fig:comparison}.

\subsection{Overview of \scname}
\label{sec:overview}

\begin{figure*}[!t]
    \centering
    \includegraphics[width=\textwidth]{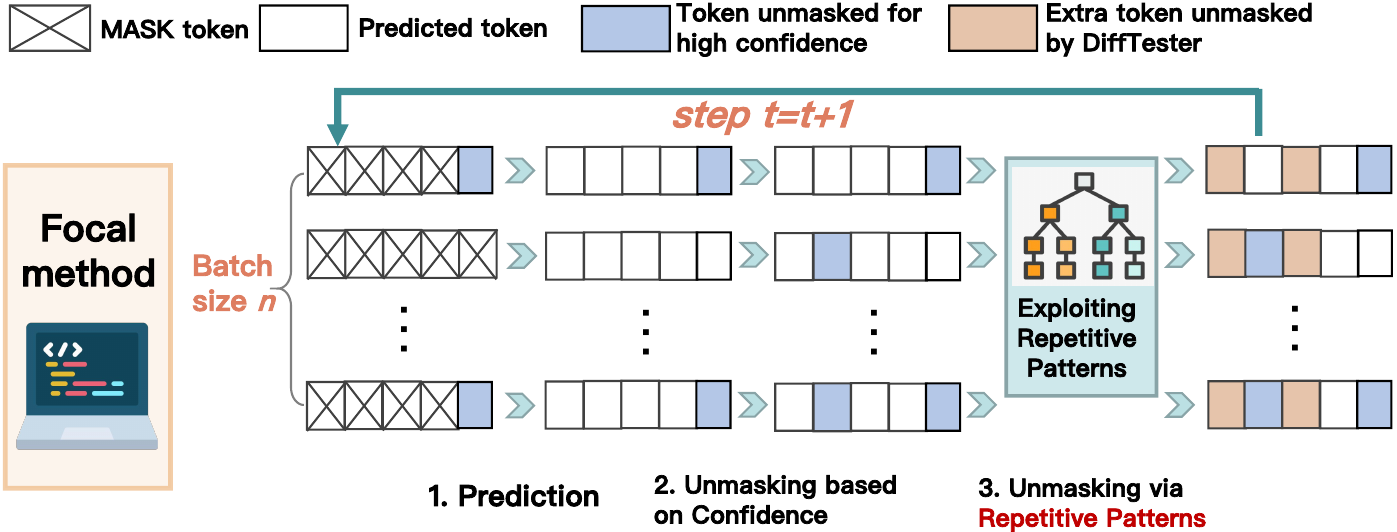}
    \caption{Overview of our proposed \name.}
    \label{fig:comparison}
\end{figure*}

We detail the overview of \name in this section.
For a given focal method, we set the batch size to $n$ (\eg $n=3, 5, 7$) and prompt the dLLM to generate one unit test case in each instance, resulting in $n$ unit test cases simultaneously. An illustration is shown in the left part of Figure~\ref{fig:comparison}

At one step $t$, for the $k$-th instance in the batch, the model first predicts tokens for all \texttt{[MASK]} positions, producing an intermediate response $\mathcal{Y}^{t,k} = (\hat{y}_i^{t,k})_{i=1}^L$.
We then follow the standard confidence based decoding defined in existing dLLMs (\eg selecting the top-$k$ tokens according to their confidence scores) to determine which tokens will be decoded~\cite{dream, dreamcoder, diffucoder}.
Subsequently and most importantly, we perform structural pattern based decoding. That's to say we extract structural patterns across the generated unit test cases and decode additional tokens corresponding to these patterns, with the detailed procedure presented in Section~\ref{sec:core}.
After completing the procedure described above, we decode all selected tokens, while the remaining tokens at \texttt{[MASK]} positions are reverted back to \texttt{[MASK]}, and the process proceeds to the next step.

\subsection{Structural Pattern Based Decoding}
\label{sec:core}

After the standard confidence based decoding, we perform structural pattern based decoding. The core is to mine structural patterns across multiple test cases. In this section, we detail how to mine these patterns through analysis of ASTs.

We begin by analyzing the key challenges involved in leveraging such structural patterns, which lie in two main aspects.

\ding{182} The first challenge is \textbf{how to flexibly and dynamically extract structural patterns across test cases}. As illustrated in the first example in Figure~\ref{fig:similarity}, although structural patterns in test cases are highly pronounced, they are flexible and diverse, making them difficult to capture only through templates.

\ding{183} The second challenge is \textbf{how to ensure sufficient diversity in the generated test cases}. Achieving high test coverage relies heavily on diversity~\cite{peacock2021automatic, yang2023improve}, yet the process of extracting shared patterns may risk reducing it. For instance, the model might generate identical input data across multiple test cases. Designing a method that can exploit shared patterns while preserving the diversity of test data is therefore another important challenge.

We next provide a detailed description of how our approach addresses these challenges and ultimately enables the effective utilization of structural patterns for acceleration.

\textbf{Extract Structural Patterns.}
Abstract Syntax Trees (ASTs) can intuitively decouple structural information from lexical information in code. Leaf nodes of an AST provide lexical information while non-leaf nodes provide structural information.
Based on this property, our approach compares the ASTs of different test cases' code and attempts to merge them as much as possible, which means identifying the nodes that are shared across different ASTs. When two or more ASTs can be merged into a non-empty tree, this indicates the existence of a structural pattern. We then locate the tokens in the intermediate test cases that correspond to the merged AST.

In practice, we find that unit test cases generated at early steps often contain many syntax errors.
Such errors propagate through the parsing process. To address this issue, we construct ASTs at the granularity of individual code lines rather than entire test cases, as illustrated in Figure~\ref{fig:comparison}.

\textbf{Ensure Sufficient Diversity.}
The diversity of unit test cases largely depends on the variability of the input data constructed for the focal method, which in turn is primarily determined by literal values such as integers or floats~\cite{yang2023improve}. To preserve this variability, we exclude the AST nodes corresponding to such literal values from the merging process. This design ensures that even if two test cases construct similar data at an intermediate step, the corresponding tokens are not retained in a single step but are instead remasked for subsequent refinement, allowing higher diversity.

After identifying the tokens that belong to the merged AST, we retain them in a single step, while the remaining tokens are remasked for refinement in the subsequent step. The details of the entire process are provided in Algorithm~\ref{alg:merge} and Algorithm~\ref{alg:method} in the Appendix.

\subsection{Additional Techniques of \scname}
\label{sec:additional}

We additionally introduce two techniques to improve the quality of the generated test cases and to further accelerate the generation process.

\ding{182} Our preliminary experiments show that directly decoding all tokens belonging to the merged AST can slightly reduce the syntactic correctness of the generated unit test cases.
We hypothesize that this is mainly because tokens with very low confidence may occasionally be retained, while in practice, we observe that such cases are extremely rare.
To address this issue, we retain only those tokens whose confidence exceeds a predefined threshold $\tau$ when guided by the merged AST, thereby ensuring that the retained tokens are more reliable for generation.

\ding{183} Although parsing code to construct ASTs is computationally inexpensive, invoking the process at every step still introduces some overhead.
Fortunately, we observe that the ASTs of code generated in consecutive steps change very little. Based on this observation, we choose not to apply \name at every step, but instead to use it intermittently after several steps, which leads to faster acceleration.
However, dLLM inference still accounts for the majority of the overall runtime. Based on a simple empirical analysis, the inference time of the dLLM is approximately 10$\times$ greater than the computational cost of the AST operations.
\section{Experiment}

\begin{figure*}[!t]
    \centering 
    \includegraphics[width=\linewidth]{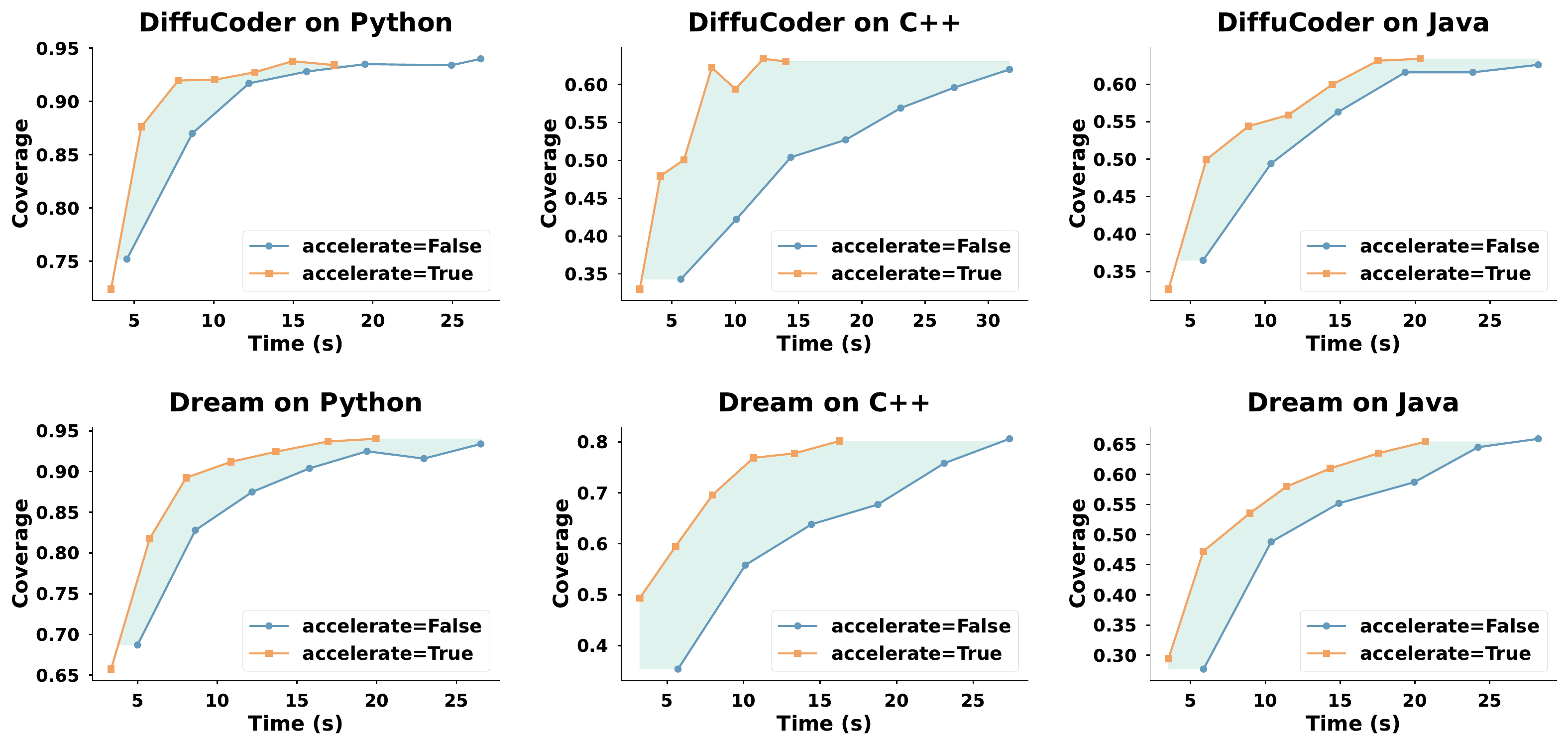}
    \caption{Comparison of line coverage with and without \name at equal decoding time.}
    \label{fig:six_results} 
\end{figure*}

In this section, we systematically evaluate the performance of \name in accelerating unit test generation. Additional experimental results including ablation studies are provided in Appendix~\ref{sec:app-exp}.

\subsection{Experiment setup}
\label{sec:setup}

\textbf{Models.}
We evaluate our approach using two representative dLLMs, namely \textsc{DiffuCoder-7B-cpGRPO}~\cite{diffucoder} and \textsc{Dream-v0-Instruct-7B}~\cite{dream}. For brevity, we refer to them as \llmname{DiffuCoder} and \llmname{Dream} in the remainder of this paper.

\textbf{Benchmarks.}
To evaluate the performance of \name, we conduct experiments on the \dataname{TestEval} benchmark~\cite{testeval}. \dataname{TestEval} is specifically designed to assess the capability of models in unit test generation, and it comprises 210 Python programs collected from LeetCode. We denote this benchmark as \dataname{TestEval-Python}.
To further validate the generalization ability of \name across different programming languages and to avoid potential data contamination, we extend the \dataname{TestEval-Python} benchmark by incorporating two additional programming languages: C++ and Java. Following the construction methodology of \dataname{TestEval-Python}, we systematically gather corresponding C++ and Java implementations of the same set of 210 programs. We denote these benchmarks as \dataname{TestEval-C++} and \dataname{TestEval-Java}, respectively.

\textbf{Evaluation Metrics.}
We adopt four widely used metrics.
\ding{182} \textit{Computational Cost (tflops)}: the average computation per batch.
\ding{183} \textit{Decoding Time (seconds, s)}: the average inference time per batch.
\ding{184} \textit{Throughput (tokens/s, tps)}: the average number of tokens generated per second per batch, excluding special tokens such as \texttt{[PAD]} and \texttt{[EOS]}.
\ding{185} \textit{Line Coverage}: the average ratio of the number of code lines covered by the generated test cases to the total number of code lines.

\textbf{Baseline.}
Since there is currently no acceleration approach specifically tailored to dLLMs for UTG, we take dLLMs without applying \name as the baseline.

\textbf{Implementation Details.}
We set the predefined generation length $L$ to 128 and the confidence threshold $\tau$ to 0.02. In addition, we apply \name once every two steps. Further implementation details are provided in Appendix~\ref{sec:hyperparameters}.

\subsection{Main Results}

We conduct evaluations on the three benchmarks and two representative models described in Section~\ref{sec:setup}.
We use different batch sizes $n$, where $n$ ranges from 1 to 7, corresponding to the number of test cases generated per focal method.
Figure~\ref{fig:six_results} presents the average test coverage achieved under varying time budgets, while Table~\ref{all-table} reports the efficiency comparison between \name and the baseline when generating the same number of test cases. And qualitative case studies are provided in Appendix~\ref{sec:case_study}.

\textbf{\name can effectively reduce the time needed to achieve a certain coverage rate.}
In the pipeline of most software development, the time available for unit testing is often limited, which makes achieving a certain coverage as fast as possible highly desirable.
As shown in Figure~\ref{fig:six_results} and Table~\ref{all-table}, to achieve the same coverage, \name can speedup up to $2\times-3\times$ compared with the baseline. For example, with a batch size $n=3$ on \dataname{TestEval-C++} using \llmname{DiffuCoder}, \name reduces the decoding time from 14.4s to 6.0s, while increasing throughput from 9.7 TPS to 23.8 TPS.
This demonstrates that the proposed \name is highly competitive and enables dLLMs to achieve higher test coverage more quickly in UTG.

\textbf{\name can effectively reduce the computational cost needed to generate test cases.}
Across nearly all settings, dLLMs achieve more than a $1.6\times$ improvement in computational cost when equipped with \name. For example, with a batch size $n=3$ on \dataname{TestEval-C++} using \llmname{DiffuCoder}, \name reduces the computational cost from 1217 TFLOPs to 430 TFLOPs. These results demonstrate that \name can substantially save computational resources.

\textbf{\name does not compromise the maximum achievable test coverage.}
The highest test coverage attainable without time constraints is another important optimization target, particularly for scenarios such as vehicle control systems~\cite{conrad2017testing, zhang2024visual}. We find that \name has little to no negative impact on the maximum test coverage and in some cases even improves it. For example, as shown in Figure~\ref{fig:six_results}, on the \dataname{TestEval-Python} benchmark with the \llmname{Dream} model, applying \name leads to a slight increase in max test coverage. We further evaluate branch coverage in Appendix~\ref{sec:branch}, where the trends are consistent with line coverage.

\textbf{\name generalizes well across different models and programming languages.}
Across various models and programming languages, \name exhibits consistent acceleration trends and similar effects on coverage. Notably, the acceleration effect of \name is more pronounced on \dataname{TestEval-C++} compared to the other two benchmarks. For example, with a batch size of 5 using \llmname{DiffuCoder}, throughput improves by 1.55× on \dataname{TestEval-Python} and 1.71× on \dataname{TestEval-Java}, whereas on \dataname{TestEval-C++} it increases by 2.45×.
We attribute this to the fact that commonly used syntactic structures in C++ exhibit greater structural similarity than those in Python and Java, which makes it easier to exploit structural patterns for acceleration.

\textbf{Comparison with an AR baseline:}
We also conducted an experiment comparing dLLM equipped with \name with an auto-regressive model for reference. Results show that dLLMs equipped with \name require less time and computational cost to achieve the same coverage in most cases compared to auto-regressive models of similar scale. The details of the experiment are provided in Appendix~\ref{sec:ar_baseline}.

\definecolor{myblue}{HTML}{d0dcf0}
\definecolor{myred}{HTML}{2f5491}

\begin{table*}[!t]
\caption{Efficiency comparison with and without \name\ across different batch sizes $n$ on \dataname{TestEval-Python}, \dataname{TestEval-C++}, and \dataname{TestEval-Java},  with \llmname{Dream} and \llmname{DiffuCoder}.}
\label{all-table}
\vspace{-0.1in}
\centering
{\Huge
\renewcommand{\arraystretch}{1.15}
\setlength{\tabcolsep}{7pt}
\resizebox{\textwidth}{!}{%
\begin{tabular}{lccccccccc}
\specialrule{3pt}{0pt}{1pt} 
\multirow{2}{*}{\textbf{Method}} & \multicolumn{3}{c}{\textbf{$n=3$}} & \multicolumn{3}{c}{\textbf{$n=5$}} & \multicolumn{3}{c}{\textbf{$n=7$}} \\
\cmidrule(lr){2-4} \cmidrule(lr){5-7} \cmidrule(lr){8-10}
& \shortstack{Computational\\Cost (tflops)} & \shortstack{Decoding\\ Time (s)} & \shortstack{Throughput\\(tps)} & \shortstack{Computational\\Cost (tflops)} & \shortstack{Decoding\\ Time (s)} & \shortstack{Throughput\\(tps)} & \shortstack{Computational\\Cost (tflops)} & \shortstack{Decoding\\ Time (s)} & \shortstack{Throughput\\(tps)} \\
\midrule
\multicolumn{10}{c}{\textbf{\dataname{TestEval-Python}}} \\
\midrule
\llmname{DiffuCoder} & 1015.59 & 12.22 & 16.97 & 1692.65 & 19.51 & 17.69 & 2369.72 & 26.78 & 18.09 \\
\cellcolor{myblue} + \name & \cellcolor{myblue}580.36 & \cellcolor{myblue}7.77 & \cellcolor{myblue}26.86 & \cellcolor{myblue}997.16 & \cellcolor{myblue}12.59 & \cellcolor{myblue}27.45 & \cellcolor{myblue}1432.00 & \cellcolor{myblue}17.57 & \cellcolor{myblue}27.42 \\
speedup             & \textcolor{myred}{\textbf{$\times1.75$}} & \textcolor{myred}{\textbf{$\times1.57$}} & \textcolor{myred}{\textbf{$\times1.58$}} & \textcolor{myred}{\textbf{$\times1.70$}} & \textcolor{myred}{\textbf{$\times1.55$}} & \textcolor{myred}{\textbf{$\times1.55$}} & \textcolor{myred}{\textbf{$\times1.65$}} & \textcolor{myred}{\textbf{$\times1.52$}} & \textcolor{myred}{\textbf{$\times1.52$}} \\

\midrule
\llmname{Dream} & 1015.59 & 12.18 & 15.61 & 1692.66 & 19.39 & 15.96 & 2369.72 & 26.53 & 16.68 \\
\cellcolor{myblue} + \name    & \cellcolor{myblue}605.05 & \cellcolor{myblue}8.04 & \cellcolor{myblue}23.59 & \cellcolor{myblue}1093.38 & \cellcolor{myblue}13.68 & \cellcolor{myblue}23.09 & \cellcolor{myblue}1644.05 & \cellcolor{myblue}19.95 & \cellcolor{myblue}22.45 \\
speedup             & \textcolor{myred}{\textbf{$\times1.68$}} & \textcolor{myred}{\textbf{$\times1.51$}} & \textcolor{myred}{\textbf{$\times1.51$}} & \textcolor{myred}{\textbf{$\times1.55$}} & \textcolor{myred}{\textbf{$\times1.42$}} & \textcolor{myred}{\textbf{$\times1.45$}} & \textcolor{myred}{\textbf{$\times1.44$}} & \textcolor{myred}{\textbf{$\times1.33$}} & \textcolor{myred}{\textbf{$\times1.35$}} \\

\midrule
\multicolumn{10}{c}{\textbf{\dataname{TestEval-C++}}} \\
\midrule

\llmname{DiffuCoder} & 1216.98 & 14.40 & 9.73 & 2098.93 & 23.08 & 9.62 & 2924.4 & 31.68 & 9.58 \\
\cellcolor{myblue} + \name     & \cellcolor{myblue}429.82 & \cellcolor{myblue}5.95 & \cellcolor{myblue}23.81 & \cellcolor{myblue}777.32 & \cellcolor{myblue}10.02 & \cellcolor{myblue}23.60 & \cellcolor{myblue}1121.37 & \cellcolor{myblue}14.00 & \cellcolor{myblue}23.20 \\
speedup             & \textcolor{myred}{\textbf{$\times2.83$}} & \textcolor{myred}{\textbf{$\times2.42$}} & \textcolor{myred}{\textbf{$\times2.45$}} & \textcolor{myred}{\textbf{$\times2.70$}} & \textcolor{myred}{\textbf{$\times2.30$}} & \textcolor{myred}{\textbf{$\times2.45$}} & \textcolor{myred}{\textbf{$\times2.61$}} & \textcolor{myred}{\textbf{$\times2.26$}} & \textcolor{myred}{\textbf{$\times2.42$}} \\

\midrule

\llmname{Dream} & 1216.98 & 14.43 & 13.11 & 2028.30 & 23.11 & 13.45 & 2839.62 & 31.60 & 13.88 \\
\cellcolor{myblue} + \name    & \cellcolor{myblue}588.17 & \cellcolor{myblue}7.96 & \cellcolor{myblue}25.93 & \cellcolor{myblue}1041.20 & \cellcolor{myblue}13.33 & \cellcolor{myblue}25.45 & \cellcolor{myblue}1614.85 & \cellcolor{myblue}19.97 & \cellcolor{myblue}24.82 \\
speedup             & \textcolor{myred}{\textbf{$\times2.07$}} & \textcolor{myred}{\textbf{$\times1.81$}} & \textcolor{myred}{\textbf{$\times1.98$}} & \textcolor{myred}{\textbf{$\times1.95$}} & \textcolor{myred}{\textbf{$\times1.73$}} & \textcolor{myred}{\textbf{$\times1.89$}} & \textcolor{myred}{\textbf{$\times1.76$}} & \textcolor{myred}{\textbf{$\times1.58$}} & \textcolor{myred}{\textbf{$\times1.79$}} \\

\midrule
\multicolumn{10}{c}{\textbf{\dataname{TestEval-Java}}} \\
\midrule

\llmname{DiffuCoder} & 1259.36 & 14.86 & 16.10 & 2098.93 & 23.85 & 16.69 & 2924.41 & 32.59 & 16.71 \\
\cellcolor{myblue} + \name     & \cellcolor{myblue}667.65 & \cellcolor{myblue}8.88 & \cellcolor{myblue}29.17 & \cellcolor{myblue}1143.00 & \cellcolor{myblue}14.46 & \cellcolor{myblue}28.58 & \cellcolor{myblue}1649.17 & \cellcolor{myblue}20.33 & \cellcolor{myblue}28.20 \\
speedup             & \textcolor{myred}{\textbf{$\times1.89$}} & \textcolor{myred}{\textbf{$\times1.67$}} & \textcolor{myred}{\textbf{$\times1.81$}} & \textcolor{myred}{\textbf{$\times1.84$}} & \textcolor{myred}{\textbf{$\times1.65$}} & \textcolor{myred}{\textbf{$\times1.71$}} & \textcolor{myred}{\textbf{$\times1.77$}} & \textcolor{myred}{\textbf{$\times1.60$}} & \textcolor{myred}{\textbf{$\times1.69$}} \\

\midrule

\llmname{Dream} & 1259.36 & 14.92 & 15.75 & 2098.93 & 24.22 & 15.79 & 2938.51 & 32.64 & 13.58 \\
\cellcolor{myblue} + \name   & \cellcolor{myblue}673.40 & \cellcolor{myblue}8.97 & \cellcolor{myblue}27.76 & \cellcolor{myblue}1130.74 & \cellcolor{myblue}14.34 & \cellcolor{myblue}28.12 & \cellcolor{myblue}1678.98 & \cellcolor{myblue}20.72 & \cellcolor{myblue}27.35 \\
speedup             & \textcolor{myred}{\textbf{$\times1.87$}} & \textcolor{myred}{\textbf{$\times1.66$}} & \textcolor{myred}{\textbf{$\times1.76$}} & \textcolor{myred}{\textbf{$\times1.86$}} & \textcolor{myred}{\textbf{$\times1.69$}} & \textcolor{myred}{\textbf{$\times1.78$}} & \textcolor{myred}{\textbf{$\times1.75$}} & \textcolor{myred}{\textbf{$\times1.58$}} & \textcolor{myred}{\textbf{$\times2.01$}} \\
\specialrule{3pt}{1pt}{0pt} 

\end{tabular}%
}
}
\vspace{-0.17in}
\end{table*}

\section{Discussion}

\subsection{Applicability to Real-World Repositories}
\definecolor{myblue}{HTML}{daede4}
\definecolor{myred}{HTML}{2a4730}
\definecolor{speedupred}{RGB}{192,0,0}
\newcommand{\speedup}[1]{{\scriptsize\textcolor{speedupred}{~($\times$#1)}}}

\begin{table}[!t]
\caption{Decoding time and line coverage on Tests4Py, comparing the baseline \llmname{DiffuCoder} and \name.}
\label{tab:tests4py}
\centering
\small
\renewcommand{\arraystretch}{1.15}
\setlength{\tabcolsep}{5pt}

\resizebox{\linewidth}{!}{
\begin{tabular}{lcccccccc}
\specialrule{1pt}{0.7pt}{0pt}

\multirow{2}{*}{\textbf{Method}}
& \multicolumn{4}{c}{\textbf{Decoding Time (s)}}
& \multicolumn{4}{c}{\textbf{Line Coverage (\%)}} \\

\cmidrule(lr){2-5}
\cmidrule(lr){6-9}

& $n=2$ & $n=3$ & $n=4$ & $n=5$
& $n=2$ & $n=3$ & $n=4$ & $n=5$ \\

\midrule

\cellcolor{gray!0} \llmname{DiffuCoder}
& \cellcolor{gray!0} 19.98
& \cellcolor{gray!0} 30.24
& \cellcolor{gray!0} 40.46
& \cellcolor{gray!0} 44.79
& \cellcolor{gray!0} 26.0
& \cellcolor{gray!0} \textbf{26.9}
& \cellcolor{gray!0} \textbf{30.0}
& \cellcolor{gray!0} 29.2 \\

\cellcolor{gray!19} + \name
& \cellcolor{gray!19} \textbf{10.27} \speedup{1.95}
& \cellcolor{gray!19} \textbf{15.86} \speedup{1.91}
& \cellcolor{gray!19} \textbf{21.18} \speedup{1.91}
& \cellcolor{gray!19} \textbf{24.18} \speedup{1.85}
& \cellcolor{gray!19} \textbf{26.3}
& \cellcolor{gray!19} 26.3
& \cellcolor{gray!19} 28.7
& \cellcolor{gray!19} \textbf{30.5} \\

\specialrule{1pt}{0.7pt}{0pt}

\end{tabular}
}
\end{table}

To examine whether \name also applies to realistic UTG scenarios, we conduct an additional evaluation on 135 samples from \textbf{Tests4Py}~\cite{smytzek2023tests4py}, which contains real bugs from popular open-source repositories such as FastAPI~\cite{Ramirez_FastAPI}, Keras~\cite{Keras_team_Keras}, and Scrapy~\cite{Scrapy_contributors_Scrapy}.
Using \llmname{DiffuCoder} as the base dLLM, we generate 2--5 unit tests for each prompt and report decoding time and line coverage in Table~\ref{tab:tests4py}.
\name achieves about $1.9\times$ speedup on average while maintaining comparable line coverage, suggesting that its structural-pattern-based acceleration can generalize beyond algorithmic benchmarks to real-world repositories.
These results suggest that the structural-pattern-based acceleration of \name is not limited to algorithmic benchmarks, but can also provide practical efficiency gains in real-world repository-level UTG scenarios without sacrificing test coverage.
More details are provided in Appendix~\ref{sec:test4py}.

\subsection{Comparison with dLLM Acceleration for General Task}

Training-free dLLM acceleration methods mainly include cache-based and sampling-based techniques, while \name belongs to the latter and remains compatible with cache-based methods.
We therefore compare \name with two representative sampling-based approaches, \textsc{EB-Sampler}~\cite{ben2025accelerated} and \textsc{SlowFast Sampling}~\cite{wei2025accelerating}, which are designed for general-purpose decoding rather than UTG. For experiments validating the compatibility between \name and cache-based methods, refer to Appendix~\ref{sec:dllm-cache}

\textsc{EB-Sampler} accelerates dLLM inference through an Entropy-Bounded unmasking procedure, which dynamically unmasks multiple tokens under a predefined approximate error tolerance.
We conduct experiments on the \dataname{TestEval-Python}, testing batch sizes of $n=3, 5, 7$, and report both average decoding time and line coverage in Table~\ref{tab:eb}.

\ding{182} Compared with \textsc{EB-Sampler}, \name achieves more effective acceleration in generating unit tests. For example, when $n=7$, \name reduces the average decoding time to 17.6s, whereas \textsc{EB-Sampler} reduces it only to 19.8s.

\ding{183} In addition, the quality of the test cases generated by \name is substantially higher. With \name, the line coverage remains almost identical to the baseline without any acceleration, while \textsc{EB-Sampler} leads to a significant drop in coverage.

\definecolor{myblue}{HTML}{daede4}
\definecolor{myred}{HTML}{2a4730}
\definecolor{speedupred}{RGB}{192,0,0}

\begin{table}[!t]
\caption{Decoding time and line coverage on \dataname{TestEval-Python} with batch sizes $n=3,5,7$, comparing the baseline \llmname{DiffuCoder}, \textsc{EB-Sampler}, and \name.}
\label{tab:eb}
\centering
\small
\renewcommand{\arraystretch}{1.15}
\setlength{\tabcolsep}{5pt} 
\resizebox{\linewidth}{!}{
\begin{tabular}{lcccccc}
\specialrule{1pt}{0.7pt}{0pt}
\multirow{2}{*}{\textbf{Method}} & \multicolumn{3}{c}{\textbf{Decoding Time (s)}} & \multicolumn{3}{c}{\textbf{Line Coverage (\%)}} \\
\cmidrule(lr){2-4} \cmidrule(lr){5-7}
& $n=3$ & $n=5$ & $n=7$ & $n=3$ & $n=5$ & $n=7$ \\
\midrule
\cellcolor{gray!0} \llmname{DiffuCoder} &
\cellcolor{gray!0} 12.2 &
\cellcolor{gray!0} 19.5 &
\cellcolor{gray!0} 26.8 &
\cellcolor{gray!0} \underline{91} &
\cellcolor{gray!0} \textbf{94} &
\cellcolor{gray!0} \underline{94} \\
\cellcolor{gray!8} + \textsc{EB-Sampler} & \cellcolor{gray!8} \underline{8.9} &
\cellcolor{gray!8} \underline{14.3} &
\cellcolor{gray!8} \underline{19.8} &
\cellcolor{gray!8} 86 &
\cellcolor{gray!8} 88 &
\cellcolor{gray!8} 88 \\
\cellcolor{gray!19} + \name         &  \cellcolor{gray!19} \textbf{7.8} \speedup{1.56}&
\cellcolor{gray!19} \textbf{12.6} \speedup{1.54} &
\cellcolor{gray!19} \textbf{17.6} \speedup{1.52}&
\cellcolor{gray!19} \textbf{92} &
\cellcolor{gray!19} \underline{93} &
\cellcolor{gray!19} \textbf{94} \\
\specialrule{1pt}{0.7pt}{0pt}
\end{tabular}
}
\end{table}

\textsc{SlowFast Sampling} is a novel dynamic sampling strategy that adaptively alternates between exploratory and accelerated decoding stages.
We conduct experiments on the \dataname{TestEval-Python}, testing batch sizes of $n=7, 10$ (for \textsc{SlowFast Sampling}) and report the average decoding time and line coverage in Table~\ref{tab:sfs}.

\ding{182} Although \textsc{SlowFast Sampling} is the fastest at the same testing batch size, it substantially degrades generation quality, achieving only 77\% line coverage even when $n=10$, underperforming even the baseline configuration with $n=3$.

\ding{183} Moreover, \textsc{SlowFast Sampling} attains identical line coverage at $n=7$ and $n=10$, indicating that its peak coverage (77\%) is significantly reduced by the acceleration strategy. In contrast, \name accelerates generation while preserving generation quality.

\definecolor{myblue}{HTML}{daede4}
\definecolor{myred}{HTML}{2a4730}
\definecolor{speedupred}{RGB}{192,0,0}

\begin{table}[!t]
\caption{Decoding time and line coverage on \dataname{TestEval-Python}, comparing the baseline \llmname{DiffuCoder}, \textsc{SlowFast Sampling}, and \name.}
\label{tab:sfs}
\centering
\small
\renewcommand{\arraystretch}{1.15}
\setlength{\tabcolsep}{5pt} 
\resizebox{\linewidth}{!}{
\begin{tabular}{lcccccc}
\specialrule{1pt}{0.7pt}{0pt}
\multirow{2}{*}{\textbf{Method}} & \multicolumn{3}{c}{\textbf{Decoding Time (s)}} & \multicolumn{3}{c}{\textbf{Line Coverage (\%)}} \\
\cmidrule(lr){2-4} \cmidrule(lr){5-7}
& $n=3$ & $n=7$ & $n=10$ & $n=3$ & $n=7$ & $n=10$ \\
\midrule
\cellcolor{gray!0} \llmname{DiffuCoder} &
\cellcolor{gray!0} 12.2 &
\cellcolor{gray!0} 26.8 &
\cellcolor{gray!0} 36.1 &
\cellcolor{gray!0} \underline{91} &
\cellcolor{gray!0} \underline{94} &
\cellcolor{gray!0} \underline{94} \\
\cellcolor{gray!8} + \textsc{SlowFast Sampling} &
\cellcolor{gray!8} \textbf{2.4} &
\cellcolor{gray!8} \textbf{5.4} &
\cellcolor{gray!8} \textbf{7.7} &
\cellcolor{gray!8} 73 &
\cellcolor{gray!8} 77 &
\cellcolor{gray!8} 77 \\
\cellcolor{gray!19} + \name &
\cellcolor{gray!19} \underline{7.8} \speedup{1.56}&
\cellcolor{gray!19} \underline{17.6} \speedup{1.52}&
\cellcolor{gray!19} \underline{24.5} \speedup{1.47}&
\cellcolor{gray!19} \textbf{92} &
\cellcolor{gray!19} \textbf{94} &
\cellcolor{gray!19} \textbf{94} \\
\specialrule{1pt}{0.7pt}{0pt}
\end{tabular}
}
\end{table}

\begin{figure*}[!ht]
    \centering
    \includegraphics[width=0.95\textwidth]{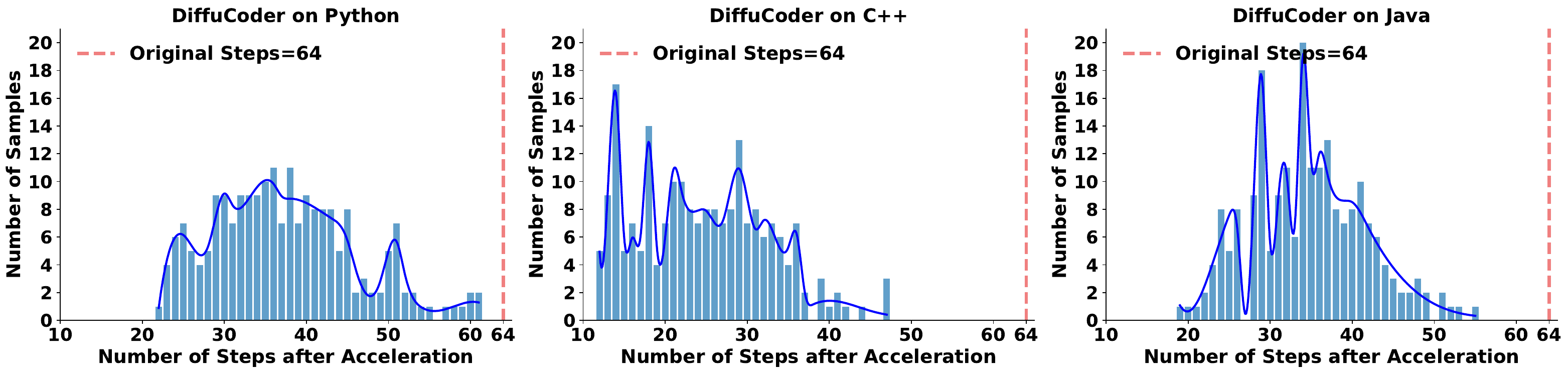}
    \caption{Distribution of the number of inference steps after applying \name on \llmname{DiffuCoder}. The red dashed line indicates the original fixed step number without acceleration.}
    \label{fig:step_distribution}
\end{figure*}

Compared to the two acceleration methods discussed earlier, \name can achieve good speedup without sacrificing generation quality. We attribute the superior performance of \name in terms of both efficiency and quality to its ability to exploit structural patterns that are inherent to the UTG task. This task-specific insight allows \name to accelerate generation without compromising the quality of the resulting test cases.

\subsection{Applying \name to Other Code Generation Tasks}

We also evaluate our method on other code generation tasks. Although these tasks may not exhibit repeated patterns as prominently as UTG, when multiple samples are generated, repeated patterns across samples can still be captured by \name and exploited to accelerate generation. For general code generation, we use the Python, C++ and Java subsets of the \textbf{HumanEval-X} benchmark~\cite{zheng2023codegeex}. For code translation, we adopt datasets from \textbf{UniTrans}~\cite{unitrans}. The experimental results are summarized in Table~\ref{tab:other_tasks}.

\definecolor{myblue}{HTML}{daede4}
\definecolor{myred}{HTML}{2a4730}
\definecolor{speedupred}{RGB}{192,0,0}

\begin{table}[!t]
\caption{Decoding time (s) and pass@1 (\%) on other code generation tasks using \llmname{DiffuCoder}.}
\label{tab:other_tasks}
\centering
\small
\renewcommand{\arraystretch}{1.15}
\setlength{\tabcolsep}{5pt} 
\resizebox{\linewidth}{!}{
\begin{tabular}{lcccccc}
\specialrule{1pt}{0.7pt}{0pt}

\multirow{2}{*}{\shortstack{\textbf{General Code}\\\textbf{Generation}}}
& \multicolumn{2}{c}{\textbf{Python}}
& \multicolumn{2}{c}{\textbf{C++}}
& \multicolumn{2}{c}{\textbf{Java}} \\

\cmidrule(lr){2-3}
\cmidrule(lr){4-5}
\cmidrule(lr){6-7}

& \textbf{Time} & \textbf{Pass@1}
& \textbf{Time} & \textbf{Pass@1}
& \textbf{Time} & \textbf{Pass@1} \\

\midrule

\cellcolor{gray!0} \llmname{DiffuCoder}
& \cellcolor{gray!0} 5.68 & \cellcolor{gray!0} \textbf{50}
& \cellcolor{gray!0} 5.73 & \cellcolor{gray!0} 11
& \cellcolor{gray!0} 6.32 & \cellcolor{gray!0} 11 \\

\cellcolor{gray!8} + \name
& \cellcolor{gray!8} \textbf{3.87} \speedup{1.47}& \cellcolor{gray!8} 49
& \cellcolor{gray!8} \textbf{4.21} \speedup{1.36}& \cellcolor{gray!8} \textbf{13}
& \cellcolor{gray!8} \textbf{5.36} \speedup{1.18}& \cellcolor{gray!8} \textbf{11} \\

\specialrule{1pt}{0.7pt}{0pt}
\end{tabular}
}

\vspace{5pt}

\resizebox{\linewidth}{!}{
\begin{tabular}{lcccccc}
\specialrule{1pt}{0.7pt}{0pt}

\multirow{2}{*}{\shortstack{\textbf{Code}\\\textbf{Translation}}}
& \multicolumn{2}{c}{\textbf{Python to C++}}
& \multicolumn{2}{c}{\textbf{C++ to Java}}
& \multicolumn{2}{c}{\textbf{Java to Python}} \\

\cmidrule(lr){2-3}
\cmidrule(lr){4-5}
\cmidrule(lr){6-7}

& \textbf{Time} & \textbf{Pass@1}
& \textbf{Time} & \textbf{Pass@1}
& \textbf{Time} & \textbf{Pass@1} \\

\midrule

\cellcolor{gray!0} \llmname{DiffuCoder}
& \cellcolor{gray!0} 11.75 & \cellcolor{gray!0} 42.9
& \cellcolor{gray!0} 11.97 & \cellcolor{gray!0} 39.0
& \cellcolor{gray!0} 12.29 & \cellcolor{gray!0} 95.8 \\

\cellcolor{gray!8} + \name
& \cellcolor{gray!8} \textbf{10.34} \speedup{1.14}& \cellcolor{gray!8} \textbf{46.2}
& \cellcolor{gray!8} \textbf{10.06} \speedup{1.19}& \cellcolor{gray!8} \textbf{40.1}
& \cellcolor{gray!8} \textbf{10.00} \speedup{1.18}& \cellcolor{gray!8} \textbf{97.5} \\

\specialrule{1pt}{0.7pt}{0pt}
\end{tabular}

}

\resizebox{\linewidth}{!}{
\begin{tabular}{lcccccc}
\specialrule{1pt}{0.7pt}{0pt}

\multirow{2}{*}{\shortstack{\textbf{Code}\\\textbf{Translation}}}
& \multicolumn{2}{c}{\textbf{C++ to Python}}
& \multicolumn{2}{c}{\textbf{Java to C++}}
& \multicolumn{2}{c}{\textbf{Python to Java}} \\

\cmidrule(lr){2-3}
\cmidrule(lr){4-5}
\cmidrule(lr){6-7}

& \textbf{Time} & \textbf{Pass@1}
& \textbf{Time} & \textbf{Pass@1}
& \textbf{Time} & \textbf{Pass@1} \\

\midrule

\cellcolor{gray!0} \llmname{DiffuCoder}
& \cellcolor{gray!0} 11.98 & \cellcolor{gray!0} \textbf{96.4}
& \cellcolor{gray!0} 12.31 & \cellcolor{gray!0} \textbf{57.0}
& \cellcolor{gray!0} 11.76 & \cellcolor{gray!0} 35.2 \\

\cellcolor{gray!8} + \name
& \cellcolor{gray!8} \textbf{9.59} \speedup{1.25}& \cellcolor{gray!8} 95.5
& \cellcolor{gray!8} \textbf{10.57} \speedup{1.16}& \cellcolor{gray!8} 56.1
& \cellcolor{gray!8} \textbf{10.26} \speedup{1.15}& \cellcolor{gray!8} \textbf{36.2} \\

\specialrule{1pt}{0.7pt}{0pt}
\end{tabular}

}
\end{table}

We conduct the experiments using \llmname{DiffuCoder} to generate 4 samples each prompt and report the average decoding time and pass@1\footnote{Average pass rate of 4 samples}.
\name achieves a modest acceleration over the baseline. The acceleration is less pronounced than in UTG because repeated structures in these code generation tasks are not as salient as those in UTG. Nevertheless, \name generally does not degrade generation quality and, in some cases, even yields higher-quality outputs.

\subsection{Distribution of Inference Steps with \scname}

In this section, we analyze the distribution of the number of inference steps after applying \name. The experiments are conducted on \llmname{DiffuCoder} across three benchmarks, with the predefined generation length set to 128. Under the default remasking strategy, the model decodes two tokens at each step, which results in 64 inference steps without \name.
The distribution of inference steps after applying \name is shown in Figure~\ref{fig:step_distribution}. We observe that the number of steps follows an approximately normal distribution, with most cases concentrated at values significantly smaller than 64. Moreover, the number of inference steps adapts to the characteristics of each focal method, making the approach well-suited for diverse testing scenarios.

\section{Conclusion}

In this paper, we present an in-depth analysis of the pervasive structural patterns in unit test generation and the characteristics of dLLMs’ inference.
Building on this analysis, we propose \name, the first acceleration framework specifically designed for dLLMs in unit test generation. The central idea of \name is to extract structural patterns in intermediate unit test cases and leverage them to decode a larger number of tokens per inference step. Extensive experimental results demonstrate that \name can substantially accelerate dLLMs in unit test generation while preserving test coverage, and that it generalizes effectively across different dLLMs and programming languages. We believe that this work provides a practical and scalable solution for improving the efficiency of unit test generation in software development.

\newpage

\section{Limitations}

Due to limited time and computational resources, we did not evaluate our method on additional models or code generation tasks. However, we believe that our approach is generally applicable to other dLLMs and code generation tasks.

\section{Ethical Considerations}

Our work focuses on improving software quality assurance and does not involve any foreseeable ethical risks. The benchmark used in our experiments is constructed entirely from publicly available sources, without the use of private or otherwise sensitive information. No human subjects, personally identifiable data, or sensitive content are involved in this research.

\section*{Acknowledgments}

We sincerely thank the anonymous reviewers for their valuable comments and constructive feedback, which helped improve this work.
We also thank the authors and developers of the models, datasets, and open-source resources used in this study for making their work publicly available and facilitating research in this area.
This work was supported by the National Natural Science Foundation of China (No. 62602364), the Beijing Natural Science Foundation (No. 4264107), and the Tsinghua University–Keystone Electrical (Zhejiang) Co., Ltd. Joint Research Center for Embodied Multimodal Artificial Intelligence (JCEMAI).

\nocite{*}
\bibliography{main.bib}

@incollection{Bengio+chapter2007,
  author    = {Bengio, Yoshua and LeCun, Yann},
  booktitle = {Large Scale Kernel Machines},
  publisher = {MIT Press},
  title     = {Scaling Learning Algorithms Towards {AI}},
  year      = {2007}
}

@article{Hinton06,
  author  = {Hinton, Geoffrey E. and Osindero, Simon and Teh, Yee Whye},
  journal = {Neural Computation},
  pages   = {1527--1554},
  title   = {A Fast Learning Algorithm for Deep Belief Nets},
  volume  = {18},
  year    = {2006}
}

@book{goodfellow2016deep,
  title     = {Deep learning},
  author    = {Goodfellow, Ian and Bengio, Yoshua and Courville, Aaron and Bengio, Yoshua},
  volume    = {1},
  year      = {2016},
  publisher = {MIT Press}
}

@inproceedings{6982627,
  author    = {Daka, Ermira and Fraser, Gordon},
  booktitle = {2014 IEEE 25th International Symposium on Software Reliability Engineering},
  title     = {A Survey on Unit Testing Practices and Problems},
  year      = {2014},
  volume    = {},
  number    = {},
  pages     = {201-211},
  doi       = {10.1109/ISSRE.2014.11}
}

@misc{testeval,
  title         = {TESTEVAL: Benchmarking Large Language Models for Test Case Generation},
  author        = {Wenhan Wang and Chenyuan Yang and Zhijie Wang and Yuheng Huang and Zhaoyang Chu and Da Song and Lingming Zhang and An Ran Chen and Lei Ma},
  year          = {2025},
  eprint        = {2406.04531},
  archiveprefix = {arXiv},
  primaryclass  = {cs.SE},
  url           = {https://arxiv.org/abs/2406.04531}
}

@misc{jain2025testgenevalrealworldunit,
  title         = {TestGenEval: A Real World Unit Test Generation and Test Completion Benchmark},
  author        = {Kush Jain and Gabriel Synnaeve and Baptiste Rozière},
  year          = {2025},
  eprint        = {2410.00752},
  archiveprefix = {arXiv},
  primaryclass  = {cs.SE},
  url           = {https://arxiv.org/abs/2410.00752}
}

@misc{liu2025longlladaunlockinglongcontext,
  title         = {LongLLaDA: Unlocking Long Context Capabilities in Diffusion LLMs},
  author        = {Xiaoran Liu and Zhigeng Liu and Zengfeng Huang and Qipeng Guo and Ziwei He and Xipeng Qiu},
  year          = {2025},
  eprint        = {2506.14429},
  archiveprefix = {arXiv},
  primaryclass  = {cs.CL},
  url           = {https://arxiv.org/abs/2506.14429}
}

@misc{nie2025scalingmaskeddiffusionmodels,
  title         = {Scaling up Masked Diffusion Models on Text},
  author        = {Shen Nie and Fengqi Zhu and Chao Du and Tianyu Pang and Qian Liu and Guangtao Zeng and Min Lin and Chongxuan Li},
  year          = {2025},
  eprint        = {2410.18514},
  archiveprefix = {arXiv},
  primaryclass  = {cs.AI},
  url           = {https://arxiv.org/abs/2410.18514}
}

@misc{song2025seeddiffusionlargescalediffusion,
  title         = {Seed Diffusion: A Large-Scale Diffusion Language Model with High-Speed Inference},
  author        = {Yuxuan Song and Zheng Zhang and Cheng Luo and Pengyang Gao and Fan Xia and Hao Luo and Zheng Li and Yuehang Yang and Hongli Yu and Xingwei Qu and Yuwei Fu and Jing Su and Ge Zhang and Wenhao Huang and Mingxuan Wang and Lin Yan and Xiaoying Jia and Jingjing Liu and Wei-Ying Ma and Ya-Qin Zhang and Yonghui Wu and Hao Zhou},
  year          = {2025},
  eprint        = {2508.02193},
  archiveprefix = {arXiv},
  primaryclass  = {cs.CL},
  url           = {https://arxiv.org/abs/2508.02193}
}

@misc{zhu2025llada15variancereducedpreference,
  title         = {LLaDA 1.5: Variance-Reduced Preference Optimization for Large Language Diffusion Models},
  author        = {Fengqi Zhu and Rongzhen Wang and Shen Nie and Xiaolu Zhang and Chunwei Wu and Jun Hu and Jun Zhou and Jianfei Chen and Yankai Lin and Ji-Rong Wen and Chongxuan Li},
  year          = {2025},
  eprint        = {2505.19223},
  archiveprefix = {arXiv},
  primaryclass  = {cs.LG},
  url           = {https://arxiv.org/abs/2505.19223}
}

@misc{prabhudesai2025diffusionbeatsautoregressivedataconstrained,
  title         = {Diffusion Beats Autoregressive in Data-Constrained Settings},
  author        = {Mihir Prabhudesai and Mengning Wu and Amir Zadeh and Katerina Fragkiadaki and Deepak Pathak},
  year          = {2025},
  eprint        = {2507.15857},
  archiveprefix = {arXiv},
  primaryclass  = {cs.LG},
  url           = {https://arxiv.org/abs/2507.15857}
}

@misc{li2025fixedtrainingfreevariablelengthdenoising,
  title         = {Beyond Fixed: Training-Free Variable-Length Denoising for Diffusion Large Language Models},
  author        = {Jinsong Li and Xiaoyi Dong and Yuhang Zang and Yuhang Cao and Jiaqi Wang and Dahua Lin},
  year          = {2025},
  eprint        = {2508.00819},
  archiveprefix = {arXiv},
  primaryclass  = {cs.CL},
  url           = {https://arxiv.org/abs/2508.00819}
}

@misc{Dreamon2025,
  title  = {DreamOn: Diffusion Language Models For Code Infilling Beyond Fixed-size Canvas},
  url    = {https://hkunlp.github.io/blog/2025/dreamon},
  author = {Wu, Zirui and Zheng, Lin and Xie, Zhihui and Ye, Jiacheng and Gao, Jiahui and Feng, Yansong and Li, Zhenguo and W., Victoria and Zhou, Guorui  and Kong, Lingpeng},
  year   = {2025}
}

@article{UT1,
  author  = {Olan, Michael},
  year    = {2003},
  month   = {01},
  pages   = {},
  title   = {Unit testing: Test early, test often},
  volume  = {19},
  journal = {Journal of Computing Sciences in Colleges - JCSC}
}

@article{UT2,
  author  = {Runeson, Per},
  year    = {2006},
  month   = {07},
  pages   = {},
  title   = {A Survey of Unit Testing Practices},
  volume  = {23},
  journal = {IEEE Software},
  doi     = {10.1109/MS.2006.91}
}

@article{search-based2,
  author  = {McMinn, Phil},
  year    = {2004},
  month   = {06},
  pages   = {105-156},
  title   = {Search-based software test data generation: a survey: Research Articles},
  volume  = {14},
  journal = {Softw. Test., Verif. Reliab.},
  doi     = {10.1002/stvr.294}
}

@unknown{symbolic_UT_1,
  author = {Baldoni, Roberto and Coppa, Emilio and D'Elia, Daniele Cono and Demetrescu, Camil and Finocchi, Irene},
  year   = {2016},
  month  = {10},
  pages  = {},
  title  = {A Survey of Symbolic Execution Techniques},
  doi    = {10.48550/arXiv.1610.00502}
}

@inproceedings{EvoSuite,
  author  = {Fraser, Gordon and Arcuri, Andrea},
  year    = {2011},
  month   = {09},
  pages   = {416-419},
  title   = {EvoSuite: Automatic test suite generation for object-oriented software},
  journal = {SIGSOFT/FSE 2011 - Proceedings of the 19th ACM SIGSOFT Symposium on Foundations of Software Engineering},
  doi     = {10.1145/2025113.2025179}
}

@unknown{Pynguin,
  author = {Lukasczyk, Stephan and Fraser, Gordon},
  year   = {2022},
  month  = {02},
  pages  = {},
  title  = {Pynguin: Automated Unit Test Generation for Python},
  doi    = {10.48550/arXiv.2202.05218}
}

@inproceedings{KLEE,
  author = {Cadar, Cristian and Dunbar, Daniel and Engler, Dawson},
  year   = {2008},
  month  = {01},
  pages  = {209-224},
  title  = {KLEE: Unassisted and Automatic Generation of High-Coverage Tests for Complex Systems Programs},
  volume = {8}
}

@article{time-consuming1,
  author  = {Barr, Earl and Harman, Mark and McMinn, Phil and Shahbaz, Muzammil and Yoo, Shin},
  year    = {2014},
  month   = {01},
  pages   = {1-1},
  title   = {The Oracle Problem in Software Testing: A Survey},
  volume  = {41},
  journal = {IEEE Transactions on Software Engineering},
  doi     = {10.1109/TSE.2014.2372785}
}

@inproceedings{NLP1,
  author = {Blasi, Arianna and Gorla, Alessandra and Ernst, Michael and Pezzè, Mauro},
  year   = {2023},
  month  = {01},
  pages  = {1-11},
  title  = {Call Me Maybe: Using NLP to Automatically Generate Unit Test Cases Respecting Temporal Constraints},
  doi    = {10.1145/3551349.3556961}
}

@article{search_base,
  author  = {Delgado-Pérez, Pedro and Ramírez, Aurora and Valle-Gómez, Kevin and Medina-Bulo, Inmaculada and Romero, José Raúl},
  year    = {2022},
  month   = {01},
  pages   = {1-17},
  title   = {InterEvo-TR: Interactive Evolutionary Test Generation With Readability Assessment},
  volume  = {PP},
  journal = {IEEE Transactions on Software Engineering},
  doi     = {10.1109/TSE.2022.3227418}
}

@inproceedings{constrained_based,
  author = {Csallner, Christoph and Tillmann, Nikolai and Smaragdakis, Yannis},
  year   = {2008},
  month  = {01},
  pages  = {281-290},
  title  = {DySy: dynamic symbolic execution for invariant inference.}
}

@inproceedings{random_based,
  author  = {Pacheco, Carlos and Lahiri, Shuvendu and Ernst, Michael and Ball, Thomas},
  year    = {2007},
  month   = {06},
  pages   = {75-84},
  title   = {Feedback-Directed Random Test Generation},
  isbn    = {0-7695-2828-7},
  journal = {Proceedings - International Conference on Software Engineering},
  doi     = {10.1109/ICSE.2007.37}
}

@article{ChatTester,
  author  = {Yuan, Zhiqiang and Liu, Mingwei and Ding, Shiji and Wang, Kaixin and Chen, Yixuan and Peng, Xin and Lou, Yiling},
  year    = {2024},
  month   = {07},
  pages   = {1703-1726},
  title   = {Evaluating and Improving ChatGPT for Unit Test Generation},
  volume  = {1},
  journal = {Proceedings of the ACM on Software Engineering},
  doi     = {10.1145/3660783}
}

@article{TestPilot,
  author  = {Schäfer, Max and Nadi, Sarah and Eghbali, Aryaz and Tip, Frank},
  year    = {2023},
  month   = {01},
  pages   = {1-21},
  title   = {An Empirical Evaluation of Using Large Language Models for Automated Unit Test Generation},
  volume  = {PP},
  journal = {IEEE Transactions on Software Engineering},
  doi     = {10.1109/TSE.2023.3334955}
}

@inproceedings{ChatUnitTest,
  author    = {Chen, Mouxiang and Liu, Zhongxin and Tao, He and Hong, Yusu and Lo, David and Xia, Xin and Sun, Jianling},
  title     = {B4: Towards Optimal Assessment of Plausible Code Solutions with Plausible Tests},
  year      = {2024},
  isbn      = {9798400712487},
  publisher = {Association for Computing Machinery},
  address   = {New York, NY, USA},
  url       = {https://doi.org/10.1145/3691620.3695536},
  doi       = {10.1145/3691620.3695536},
  pages     = {1693–1705},
  numpages  = {13},
  location  = {Sacramento, CA, USA},
  series    = {ASE '24}
}

@inproceedings{gren2017relation,
  title        = {On the relation between unit testing and code quality},
  author       = {Gren, Lucas and Antinyan, Vard},
  booktitle    = {2017 43rd Euromicro Conference on Software Engineering and Advanced Applications (SEAA)},
  pages        = {52--56},
  year         = {2017},
  organization = {IEEE}
}

@article{shang2025large,
  title     = {A large-scale empirical study on fine-tuning large language models for unit testing},
  author    = {Shang, Ye and Zhang, Quanjun and Fang, Chunrong and Gu, Siqi and Zhou, Jianyi and Chen, Zhenyu},
  journal   = {Proceedings of the ACM on Software Engineering},
  volume    = {2},
  number    = {ISSTA},
  pages     = {1678--1700},
  year      = {2025},
  publisher = {ACM New York, NY, USA}
}

@article{chipounov2011s2e,
  title     = {S2E: A platform for in-vivo multi-path analysis of software systems},
  author    = {Chipounov, Vitaly and Kuznetsov, Volodymyr and Candea, George},
  journal   = {Acm Sigplan Notices},
  volume    = {46},
  number    = {3},
  pages     = {265--278},
  year      = {2011},
  publisher = {ACM New York, NY, USA}
}

@article{tufano2020unit,
  title   = {Unit test case generation with transformers and focal context},
  author  = {Tufano, Michele and Drain, Dawn and Svyatkovskiy, Alexey and Deng, Shao Kun and Sundaresan, Neel},
  journal = {arXiv preprint arXiv:2009.05617},
  year    = {2020}
}

@article{wang2024software,
  title     = {Software testing with large language models: Survey, landscape, and vision},
  author    = {Wang, Junjie and Huang, Yuchao and Chen, Chunyang and Liu, Zhe and Wang, Song and Wang, Qing},
  journal   = {IEEE Transactions on Software Engineering},
  volume    = {50},
  number    = {4},
  pages     = {911--936},
  year      = {2024},
  publisher = {IEEE}
}

@inproceedings{bhatia2024unit,
  title     = {Unit test generation using generative AI: A comparative performance analysis of autogeneration tools},
  author    = {Bhatia, Shreya and Gandhi, Tarushi and Kumar, Dhruv and Jalote, Pankaj},
  booktitle = {Proceedings of the 1st International Workshop on Large Language Models for Code},
  pages     = {54--61},
  year      = {2024}
}

@inproceedings{robinson2011scaling,
  title        = {Scaling up automated test generation: Automatically generating maintainable regression unit tests for programs},
  author       = {Robinson, Brian and Ernst, Michael D and Perkins, Jeff H and Augustine, Vinay and Li, Nuo},
  booktitle    = {2011 26th IEEE/ACM International Conference on Automated Software Engineering (ASE 2011)},
  pages        = {23--32},
  year         = {2011},
  organization = {IEEE}
}

@inproceedings{li2006large,
  title     = {Large-scale software unit testing on the grid.},
  author    = {Li, Yaohang and Dong, Tao and Zhang, Xinyu and Song, Yong-duan and Yuan, Xiaohong},
  booktitle = {GrC},
  pages     = {596--599},
  year      = {2006}
}

@article{hui2024qwen2,
  title   = {Qwen2. 5-coder technical report},
  author  = {Hui, Binyuan and Yang, Jian and Cui, Zeyu and Yang, Jiaxi and Liu, Dayiheng and Zhang, Lei and Liu, Tianyu and Zhang, Jiajun and Yu, Bowen and Lu, Keming and others},
  journal = {arXiv preprint arXiv:2409.12186},
  year    = {2024}
}

@article{achiam2023gpt,
  title   = {Gpt-4 technical report},
  author  = {Achiam, Josh and Adler, Steven and Agarwal, Sandhini and Ahmad, Lama and Akkaya, Ilge and Aleman, Florencia Leoni and Almeida, Diogo and Altenschmidt, Janko and Altman, Sam and Anadkat, Shyamal and others},
  journal = {arXiv preprint arXiv:2303.08774},
  year    = {2023}
}

@inproceedings{yang2024evaluation,
  title     = {On the evaluation of large language models in unit test generation},
  author    = {Yang, Lin and Yang, Chen and Gao, Shutao and Wang, Weijing and Wang, Bo and Zhu, Qihao and Chu, Xiao and Zhou, Jianyi and Liang, Guangtai and Wang, Qianxiang and others},
  booktitle = {Proceedings of the 39th IEEE/ACM International Conference on Automated Software Engineering},
  pages     = {1607--1619},
  year      = {2024}
}

@article{dream,
  title   = {Dream 7b: Diffusion large language models},
  author  = {Ye, Jiacheng and Xie, Zhihui and Zheng, Lin and Gao, Jiahui and Wu, Zirui and Jiang, Xin and Li, Zhenguo and Kong, Lingpeng},
  journal = {arXiv preprint arXiv:2508.15487},
  year    = {2025}
}

@article{diffucoder,
  title   = {DiffuCoder: Understanding and Improving Masked Diffusion Models for Code Generation},
  author  = {Gong, Shansan and Zhang, Ruixiang and Zheng, Huangjie and Gu, Jiatao and Jaitly, Navdeep and Kong, Lingpeng and Zhang, Yizhe},
  journal = {arXiv preprint arXiv:2506.20639},
  year    = {2025}
}

@article{mercury,
  title   = {Mercury: Ultra-fast language models based on diffusion},
  author  = {Khanna, Samar and Kharbanda, Siddhant and Li, Shufan and Varma, Harshit and Wang, Eric and Birnbaum, Sawyer and Luo, Ziyang and Miraoui, Yanis and Palrecha, Akash and Ermon, Stefano and others},
  journal = {arXiv preprint arXiv:2506.17298},
  year    = {2025}
}

@article{llada,
  title   = {Large language diffusion models},
  author  = {Nie, Shen and Zhu, Fengqi and You, Zebin and Zhang, Xiaolu and Ou, Jingyang and Hu, Jun and Zhou, Jun and Lin, Yankai and Wen, Ji-Rong and Li, Chongxuan},
  journal = {arXiv preprint arXiv:2502.09992},
  year    = {2025}
}

@misc{dreamcoder,
  title  = {Dream-Coder 7B},
  url    = {https://hkunlp.github.io/blog/2025/dream-coder},
  author = {Xie, Zhihui and Ye, Jiacheng and Zheng, Lin and Gao, Jiahui and Dong, Jingwei and Wu, Zirui and Zhao, Xueliang and Gong, Shansan and Jiang, Xin and Li, Zhenguo and Kong, Lingpeng},
  year   = {2025}
}

@misc{gemini_diffusion,
  title  = {Gemini Diffusion},
  author = {Google DeepMind},
  year   = {2025},
  url    = {https://deepmind.google/models/gemini-diffusion/}
}

@unknown{Fast-dLLM,
  author = {Wu, Chengyue and Zhang, Hao and Xue, Shuchen and Liu, Zhijian and Diao, Shizhe and Zhu, Ligeng and Luo, Ping and Han, Song and Xie, Enze},
  year   = {2025},
  month  = {05},
  pages  = {},
  title  = {Fast-dLLM: Training-free Acceleration of Diffusion LLM by Enabling KV Cache and Parallel Decoding},
  doi    = {10.48550/arXiv.2505.22618}
}

@unknown{dLLM-Cache,
  author = {Liu, Zhiyuan and Yang, Yicun and Zhang, Yaojie and Chen, Junjie and Zou, Chang and Wei, Qingyan and Wang, Shaobo and Zhang, Linfeng},
  year   = {2025},
  month  = {05},
  pages  = {},
  title  = {dLLM-Cache: Accelerating Diffusion Large Language Models with Adaptive Caching},
  doi    = {10.13140/RG.2.2.30694.13127}
}

@inproceedings{blockdiffusion,
  title     = {Block Diffusion: Interpolating Between Autoregressive and Diffusion Language Models},
  author    = {Marianne Arriola and Subham Sekhar Sahoo and Aaron Gokaslan and Zhihan Yang and Zhixuan Qi and Jiaqi Han and Justin T Chiu and Volodymyr Kuleshov},
  booktitle = {The Thirteenth International Conference on Learning Representations},
  year      = {2025},
  url       = {https://openreview.net/forum?id=tyEyYT267x}
}

@article{zhang2025beyond,
  title   = {Beyond Autoregression: An Empirical Study of Diffusion Large Language Models for Code Generation},
  author  = {Li, Chengze and Zhang, Yitong and Li, Jia and Cai, Liyi and Li, Ge and others},
  journal = {arXiv preprint arXiv:2509.11252},
  year    = {2025}
}

@article{zeng2025treediff,
  title   = {TreeDiff: AST-Guided Code Generation with Diffusion LLMs},
  author  = {Zeng, Yiming and Cao, Jinghan and Li, Zexin and Chen, Yiming and Ren, Tao and Xiang, Dawei and Wu, Xidong and Gao, Shangqian and Yu, Tingting},
  journal = {arXiv preprint arXiv:2508.01473},
  year    = {2025}
}

@article{mundler2025constrained,
  title   = {Constrained Decoding of Diffusion LLMs with Context-Free Grammars},
  author  = {M{\"u}ndler, Niels and Dekoninck, Jasper and Vechev, Martin},
  journal = {arXiv preprint arXiv:2508.10111},
  year    = {2025}
}

@inproceedings{peacock2021automatic,
  title        = {Automatic equivalent mutants classification using abstract syntax tree neural networks},
  author       = {Peacock, Samuel and Deng, Lin and Dehlinger, Josh and Chakraborty, Suranjan},
  booktitle    = {2021 IEEE International Conference on Software Testing, Verification and Validation Workshops (ICSTW)},
  pages        = {13--18},
  year         = {2021},
  organization = {IEEE}
}

@article{deng2023investigating,
  title   = {Investigating data contamination in modern benchmarks for large language models},
  author  = {Deng, Chunyuan and Zhao, Yilun and Tang, Xiangru and Gerstein, Mark and Cohan, Arman},
  journal = {arXiv preprint arXiv:2311.09783},
  year    = {2023}
}

@article{wei2025accelerating,
  title   = {Accelerating Diffusion Large Language Models with SlowFast: The Three Golden Principles},
  author  = {Wei, Qingyan and Zhang, Yaojie and Liu, Zhiyuan and Liu, Dongrui and Zhang, Linfeng},
  journal = {arXiv preprint arXiv:2506.10848},
  year    = {2025}
}

@inproceedings{lemieux2023codamosa,
  title        = {Codamosa: Escaping coverage plateaus in test generation with pre-trained large language models},
  author       = {Lemieux, Caroline and Inala, Jeevana Priya and Lahiri, Shuvendu K and Sen, Siddhartha},
  booktitle    = {2023 IEEE/ACM 45th International Conference on Software Engineering (ICSE)},
  pages        = {919--931},
  year         = {2023},
  organization = {IEEE}
}

@article{israel2025accelerating,
  title   = {Accelerating Diffusion LLMs via Adaptive Parallel Decoding},
  author  = {Israel, Daniel and Broeck, Guy Van den and Grover, Aditya},
  journal = {arXiv preprint arXiv:2506.00413},
  year    = {2025}
}

@article{zamprogno2022dynamic,
  title     = {Dynamic human-in-the-loop assertion generation},
  author    = {Zamprogno, Lucas and Hall, Braxton and Holmes, Reid and Atlee, Joanne M},
  journal   = {IEEE Transactions on Software Engineering},
  volume    = {49},
  number    = {4},
  pages     = {2337--2351},
  year      = {2022},
  publisher = {IEEE}
}

@article{yang2023improve,
  title     = {Improve model testing by integrating bounded model checking and coverage guided fuzzing},
  author    = {Yang, Yixiao},
  journal   = {Electronics},
  volume    = {12},
  number    = {7},
  pages     = {1573},
  year      = {2023},
  publisher = {MDPI}
}

@article{sun2023abstract,
  title   = {Abstract syntax tree for programming language understanding and representation: How far are we?},
  author  = {Sun, Weisong and Fang, Chunrong and Miao, Yun and You, Yudu and Yuan, Mengzhe and Chen, Yuchen and Zhang, Quanjun and Guo, An and Chen, Xiang and Liu, Yang and others},
  journal = {arXiv preprint arXiv:2312.00413},
  year    = {2023}
}

@inproceedings{suttichaya2022source,
  title        = {Source code plagiarism detection based on abstract syntax tree fingerprintings},
  author       = {Suttichaya, Vasin and Eakvorachai, Niracha and Lurkraisit, Tunchanok},
  booktitle    = {2022 17th International Joint Symposium on Artificial Intelligence and Natural Language Processing (iSAI-NLP)},
  pages        = {1--6},
  year         = {2022},
  organization = {IEEE}
}

@article{dakhel2024effective,
  title     = {Effective test generation using pre-trained large language models and mutation testing},
  author    = {Dakhel, Arghavan Moradi and Nikanjam, Amin and Majdinasab, Vahid and Khomh, Foutse and Desmarais, Michel C},
  journal   = {Information and Software Technology},
  volume    = {171},
  pages     = {107468},
  year      = {2024},
  publisher = {Elsevier}
}

@book{aniche2022effective,
  title     = {Effective Software Testing: A developer's guide},
  author    = {Aniche, Maur{\'\i}cio},
  year      = {2022},
  publisher = {Simon and Schuster}
}

@article{yang2025mmada,
  title   = {Mmada: Multimodal large diffusion language models},
  author  = {Yang, Ling and Tian, Ye and Li, Bowen and Zhang, Xinchen and Shen, Ke and Tong, Yunhai and Wang, Mengdi},
  journal = {arXiv preprint arXiv:2505.15809},
  year    = {2025}
}

@article{you2025llada,
  title   = {Llada-v: Large language diffusion models with visual instruction tuning},
  author  = {You, Zebin and Nie, Shen and Zhang, Xiaolu and Hu, Jun and Zhou, Jun and Lu, Zhiwu and Wen, Ji-Rong and Li, Chongxuan},
  journal = {arXiv preprint arXiv:2505.16933},
  year    = {2025}
}

@article{song2025sparse,
  title   = {Sparse-dLLM: Accelerating Diffusion LLMs with Dynamic Cache Eviction},
  author  = {Song, Yuerong and Liu, Xiaoran and Li, Ruixiao and Liu, Zhigeng and Huang, Zengfeng and Guo, Qipeng and He, Ziwei and Qiu, Xipeng},
  journal = {arXiv preprint arXiv:2508.02558},
  year    = {2025}
}

@article{sahoo2024simple,
  title   = {Simple and effective masked diffusion language models},
  author  = {Sahoo, Subham and Arriola, Marianne and Schiff, Yair and Gokaslan, Aaron and Marroquin, Edgar and Chiu, Justin and Rush, Alexander and Kuleshov, Volodymyr},
  journal = {Advances in Neural Information Processing Systems},
  volume  = {37},
  pages   = {130136--130184},
  year    = {2024}
}

@article{discreteVLA,
  title   = {Discrete Diffusion VLA: Bringing Discrete Diffusion to Action Decoding in Vision-Language-Action Policies},
  author  = {Liang, Zhixuan and Li, Yizhuo and Yang, Tianshuo and Wu, Chengyue and Mao, Sitong and Pei, Liuao and Yang, Xiaokang and Pang, Jiangmiao and Mu, Yao and Luo, Ping},
  journal = {arXiv preprint arXiv:2508.20072},
  year    = {2025}
}

@article{li2025survey,
  title   = {A survey on diffusion language models},
  author  = {Li, Tianyi and Chen, Mingda and Guo, Bowei and Shen, Zhiqiang},
  journal = {arXiv preprint arXiv:2508.10875},
  year    = {2025}
}

@unknown{Any-Order,
  author = {Kim, Jaeyeon and Cheuk-Kit, Lee and Domingo-Enrich, Carles and Du, Yilun and Kakade, Sham and Ngotiaoco, Timothy and Chen, Sitan and Albergo, Michael},
  year   = {2025},
  month  = {08},
  pages  = {},
  title  = {Any-Order Flexible Length Masked Diffusion},
  doi    = {10.48550/arXiv.2509.01025}
}

@unknown{Edit-Flows,
  author = {Havasi, Marton and Karrer, Brian and Gat, Itai and Chen, Ricky},
  year   = {2025},
  month  = {06},
  pages  = {},
  title  = {Edit Flows: Flow Matching with Edit Operations},
  doi    = {10.48550/arXiv.2506.09018}
}

@inproceedings{kim2025train,
  title     = {Train for the Worst, Plan for the Best: Understanding Token Ordering in Masked Diffusions},
  author    = {Jaeyeon Kim and Kulin Shah and Vasilis Kontonis and Sham M. Kakade and Sitan Chen},
  booktitle = {Forty-second International Conference on Machine Learning},
  year      = {2025},
  url       = {https://openreview.net/forum?id=DjJmre5IkP}
}

@inproceedings{zhao2025d,
  title     = {d1: Scaling Reasoning in Diffusion Large Language Models via Reinforcement Learning},
  author    = {Siyan Zhao and Devaansh Gupta and Qinqing Zheng and Aditya Grover},
  booktitle = {ES-FoMo III: 3rd Workshop on Efficient Systems for Foundation Models},
  year      = {2025},
  url       = {https://openreview.net/forum?id=t8oYNHAvM9}
}

@article{zhang2024visual,
  title   = {Visual adversarial attack on vision-language models for autonomous driving},
  author  = {Zhang, Tianyuan and Wang, Lu and Zhang, Xinwei and Zhang, Yitong and Jia, Boyi and Liang, Siyuan and Hu, Shengshan and Fu, Qiang and Liu, Aishan and Liu, Xianglong},
  journal = {arXiv preprint arXiv:2411.18275},
  year    = {2024}
}

@incollection{conrad2017testing,
  title     = {Testing automotive control software},
  author    = {Conrad, Mirko and Fey, Ines},
  booktitle = {Automotive Embedded Systems Handbook},
  pages     = {11--1},
  year      = {2017},
  publisher = {CRC Press}
}

@article{ben2025accelerated,
  title   = {Accelerated Sampling from Masked Diffusion Models via Entropy Bounded Unmasking},
  author  = {Ben-Hamu, Heli and Gat, Itai and Severo, Daniel and Nolte, Niklas and Karrer, Brian},
  journal = {arXiv preprint arXiv:2505.24857},
  year    = {2025}
}

@misc{chen2023acceleratinglargelanguagemodel,
  title         = {Accelerating Large Language Model Decoding with Speculative Sampling},
  author        = {Charlie Chen and Sebastian Borgeaud and Geoffrey Irving and Jean-Baptiste Lespiau and Laurent Sifre and John Jumper},
  year          = {2023},
  eprint        = {2302.01318},
  archiveprefix = {arXiv},
  primaryclass  = {cs.CL},
  url           = {https://arxiv.org/abs/2302.01318}
}

@misc{leviathan2023fastinferencetransformersspeculative,
  title         = {Fast Inference from Transformers via Speculative Decoding},
  author        = {Yaniv Leviathan and Matan Kalman and Yossi Matias},
  year          = {2023},
  eprint        = {2211.17192},
  archiveprefix = {arXiv},
  primaryclass  = {cs.LG},
  url           = {https://arxiv.org/abs/2211.17192}
}

@article{unitrans,
  author     = {Yang, Zhen and Liu, Fang and Yu, Zhongxing and Keung, Jacky Wai and Li, Jia and Liu, Shuo and Hong, Yifan and Ma, Xiaoxue and Jin, Zhi and Li, Ge},
  title      = {Exploring and Unleashing the Power of Large Language Models in Automated Code Translation},
  year       = {2024},
  issue_date = {July 2024},
  publisher  = {Association for Computing Machinery},
  address    = {New York, NY, USA},
  volume     = {1},
  number     = {FSE},
  url        = {https://doi.org/10.1145/3660778},
  doi        = {10.1145/3660778},
  journal    = {Proc. ACM Softw. Eng.},
  month      = jul,
  articleno  = {71},
  numpages   = {24}
}

@inproceedings{zheng2023codegeex,
  title     = {CodeGeeX: A Pre-Trained Model for Code Generation with Multilingual Benchmarking on HumanEval-X},
  author    = {Qinkai Zheng and Xiao Xia and Xu Zou and Yuxiao Dong and Shan Wang and Yufei Xue and Zihan Wang and Lei Shen and Andi Wang and Yang Li and Teng Su and Zhilin Yang and Jie Tang},
  booktitle = {Proceedings of the 29th ACM SIGKDD Conference on Knowledge Discovery and Data Mining},
  pages     = {5673--5684},
  year      = {2023}
}

@misc{sdar,
  title         = {SDAR: A Synergistic Diffusion-AutoRegression Paradigm for Scalable Sequence Generation},
  author        = {Shuang Cheng and Yihan Bian and Dawei Liu and Linfeng Zhang and Qian Yao and Zhongbo Tian and Wenhai Wang and Qipeng Guo and Kai Chen and Biqing Qi and Bowen Zhou},
  year          = {2025},
  eprint        = {2510.06303},
  archiveprefix = {arXiv},
  primaryclass  = {cs.LG},
  url           = {https://arxiv.org/abs/2510.06303}
}

@misc{smytzek2023tests4py,
  title         = {Tests4Py: A Benchmark for System Testing},
  author        = {Marius Smytzek and Martin Eberlein and Batuhan Serce and Lars Grunske and Andreas Zeller},
  year          = {2023},
  eprint        = {2307.05147},
  archiveprefix = {arXiv},
  primaryclass  = {cs.SE}
}

@software{Ramirez_FastAPI,
  author  = {Ramírez, Sebastián},
  license = {MIT},
  title   = {{FastAPI}},
  url     = {https://github.com/fastapi/fastapi}
}

@software{Scrapy_contributors_Scrapy,
  author = {{Scrapy contributors}},
  title  = {{Scrapy}},
  url    = {https://scrapy.org}
}

@software{Keras_team_Keras,
  author = {{Keras team}},
  title  = {{Keras}},
  url    = {https://keras.io/}
}

@misc{zhang2026lookaheadthenverifyreliableconstraineddecoding,
  title         = {Lookahead-then-Verify: Reliable Constrained Decoding for Diffusion LLMs under Context-Free Grammars},
  author        = {Yitong Zhang and Yongmin Li and Yuetong Liu and Jia Li and Xiaoran Jia and Zherui Li and Ge Li},
  year          = {2026},
  eprint        = {2602.00612},
  archiveprefix = {arXiv},
  primaryclass  = {cs.CL},
  url           = {https://arxiv.org/abs/2602.00612}
}

@misc{chen2026coactactionpreservingobservationcompression,
  title         = {CoACT: Action-Preserving Observation Compression for Coding Agents},
  author        = {Haorui Chen and Yuancheng Zhu and Yitong Zhang and Jia Li},
  year          = {2026},
  eprint        = {2607.02911},
  archiveprefix = {arXiv},
  primaryclass  = {cs.SE},
  url           = {https://arxiv.org/abs/2607.02911}
}

@misc{li2026diffuguardintrinsicsafetylost,
  title         = {DiffuGuard: How Intrinsic Safety is Lost and Found in Diffusion Large Language Models},
  author        = {Zherui Li and Zheng Nie and Zhenhong Zhou and Yue Liu and Yitong Zhang and Yu Cheng and Qingsong Wen and Kun Wang and Yufei Guo and Jiaheng Zhang},
  year          = {2026},
  eprint        = {2509.24296},
  archiveprefix = {arXiv},
  primaryclass  = {cs.CL},
  url           = {https://arxiv.org/abs/2509.24296}
}

@misc{yang2026improvingsamplingmaskeddiffusion,
  title         = {Improving Sampling for Masked Diffusion Models via Information Gain},
  author        = {Kaisen Yang and Jayden Teoh and Kaicheng Yang and Yitong Zhang and Alex Lamb},
  year          = {2026},
  eprint        = {2602.18176},
  archiveprefix = {arXiv},
  primaryclass  = {cs.CL},
  url           = {https://arxiv.org/abs/2602.18176}
}

\newpage
\appendix
\section{LLM Usage Statement}

We used Large Language Models solely as a tool to assist with the linguistic aspects of our manuscript. Specifically, the model was employed to help with translation, refine grammar, enhance clarity, improve conciseness, and optimize word choice. At no point did the model contribute any new ideas, original content, or substantive changes to the manuscript. Its function was limited strictly to editing and language optimization, ensuring that the scientific integrity and originality of our work remained entirely unaffected by the use of the model.

\section{Reproducibility statement}

To ensure the reproducibility of our results, we have made all relevant code and datasets publicly available at \url{https://github.com/THU-Agent/DiffuTester}. The repository includes detailed instructions for setting up the environment, running experiments, and reproducing all results presented in the paper.  We encourage the community to use these resources to verify our findings and to facilitate further research in this area.

All existing artifacts used in this work, including models, benchmarks, and source code repositories, were utilized in a manner consistent with their original licenses and intended research purposes. Our extended benchmark is constructed entirely from publicly available sources and is released solely for research and evaluation purposes. The source code and benchmark of this project are released under the MIT License.

\section{Additional Discussion}

\subsection{Detailed Setup}

The results reported in Table~\ref{all-table} were collected on a server with 8 NVIDIA A100-SXM4-40G GPUs (PCIe between CPU and GPU), and two Intel Xeon Gold 6348 CPUs, with 28 cores per socket and 56 physical cores in total.

\subsubsection{Models}

We primarily conducted experiments using the dLLMs \llmname{DiffuCoder-7B-cpGRPO}~\citep{diffucoder} (\llmname{DiffuCoder} in this paper for short) and \llmname{Dream-v0-Instruct-7B}~\cite{dream} (\llmname{Dream} in this paper for short).

\textbf{DiffuCoder-7B-cpGRPO} is a code generation dLLM developed by Apple. It is a refined variant of DiffuCoder-Instruct, further improved using Coupled-GRPO reinforcement learning. The model is initialized from DiffuCoder-7B-Instruct and post-trained on 21,000 code samples for one epoch.

\textbf{Dream-v0-Instruct-7B} is a variant of the Dream 7B model developed by the HKU NLP Group with 7 billion parameters. It is an instruction-tuned (SFT) large diffusion language model. It is trained on a mixture of text, math, and code data, leveraging weight initialization from auto-regressive models for more efficient learning, and it demonstrates strong performance on general, coding, and reasoning tasks.

\subsubsection{Metrics}

The following provides a detailed description of the metrics measured in our main experiments.

\textbf{Coverage.} A software testing metric indicating the percentage of source code lines (or branches) exercised by the generated test cases. Higher coverage generally means that the generated tests explore more of the code base, leading to better test quality. In our experiments, we use line coverage, calculated as the number of code lines covered by the test cases divided by the total number of code lines.

\textbf{Computational Cost (tflops)}. The overall amount of computation required by the model during inference reflects the hardware resources needed to complete the task. In our experiments, we measure the average computational cost to generate a certain number of test cases consumed per problem across the entire benchmark.

\textbf{Time (s).} Time needed during inference to generate a certain number of test cases. In our experiments, we measure the average inference time consumed per problem across the entire benchmark.

\textbf{Throughput (tps).} The number of tokens generated per second (tokens per second, tps). It reflects how many outputs can be produced in parallel and is critical for large-scale deployment. In our experiments, throughput is calculated as the total number of tokens generated across the entire benchmark (excluding special tokens such as \texttt{[EOS]} and \texttt{[PAD]}) divided by the total GPU time used.

\subsubsection{Hyperparameters} \label{sec:hyperparameters}

Hyperparameters we used in the main experiments are listed in Table~\ref{tab:hyperparams}:

\begin{table*}[ht]
\centering
\small
\caption{Hyperparameters used in the main experiments.}
\resizebox{\textwidth}{!}{
\begin{tabular}{llcccccc}
\toprule
\textbf{Model} & \textbf{Language} & \textbf{Steps} & \textbf{Temperature} & \textbf{Max length} & \textbf{Alg. temp.}\footnote{Confidence as logit, apply softmax, and sample positions to unmask.} & \textbf{Threshold}\footnote{Only tokens whose confidence is above the threshold are unmasked.} & \textbf{Step size}\footnote{Acceleration algorithm applied every how many steps.} \\
\midrule
\multirow{3}{*}{DiffuCoder}
  & Python & 64 & 1.5 & 128 & 0.2 & 0.02 & 2 \\
  & C++    & 64 & 1.5 & 192 & 0.2 & 0.02 & 2 \\
  & Java   & 64 & 1.5 & 192 & 0.2 & 0.02 & 2 \\
\midrule
\multirow{3}{*}{Dream}
  & Python & 64 & 1.0 & 128 & 0.2 & 0.02 & 2 \\
  & C++    & 64 & 1.0 & 192 & 0.2 & 0.02 & 2 \\
  & Java   & 64 & 1.0 & 192 & 0.2 & 0.02 & 2 \\
\bottomrule
\end{tabular}
}
\label{tab:hyperparams}
\end{table*}

In our experiments with the \textsc{EB-Sampler} method, we adopted the parameters recommended in the original paper~\cite{ben2025accelerated}. We used confidence as the \textit{error proxy function}. And we used $\gamma=0.1$.

In our experiments with the \textsc{SlowFast Sampling}~\cite{wei2025accelerating} method, we adopted default parameters in the implementation by the authors.\footnote{https://github.com/LiangrunFlora/Slow-Fast-Sampling/blob/main/slow-fast-sampling/sampling\_utils/dream\_generation\_utils.py}

\subsubsection{Benchmark}

To evaluate the performance of our method, we experiment on the TestEval benchmark~\cite{testeval}. TestEval is a benchmark for test case generation with LLMs. It includes 210 Python programs from an online programming platform, LeetCode. It has three metrics: overall coverage, targeted line/branch coverage, and targeted path coverage.

To further assess the generalizability of our method across multiple programming languages, we extend the TestEval benchmark by collecting corresponding C++ and Java implementations for the same set of 210 problems. These additional solutions are obtained from a publicly available GitHub repository licensed under the MIT license (see https://github.com/walkccc/LeetCode for details).

Building on TestEval, we also adapted the evaluation code for line coverage and branch coverage in Java and C++. For coverage measurement, we used pytest for Python, Maven for Java, and gcov for C++.

\subsection{More Experiments}
\label{sec:app-exp}

\subsubsection{Details of the experiment on Tests4Py}
\label{sec:test4py}

In this section, we describe the evaluation pipeline for our experiments on Tests4Py.

We first collect the source code of all bugs included in Tests4Py\footnote{\url{https://github.com/smythi93/Tests4Py}} and, for simplicity, retain only bugs localized within a single function. We then build the execution environment for each bug in each repository. Bugs for which the environment cannot be successfully built are discarded, resulting in a final benchmark of 135 instances.

We then construct prompts using the code of the target function together with surrounding contextual code from the same file. The hyperparameters used in this experiment are identical to those used for \dataname{TestEval-Python}.

\subsubsection{Comparison with AR model}
\label{sec:ar_baseline}

In this section, we present a comparison between our method and an AR baseline model.

\begin{figure*}[!ht]
    \centering
    \includegraphics[width=\textwidth]{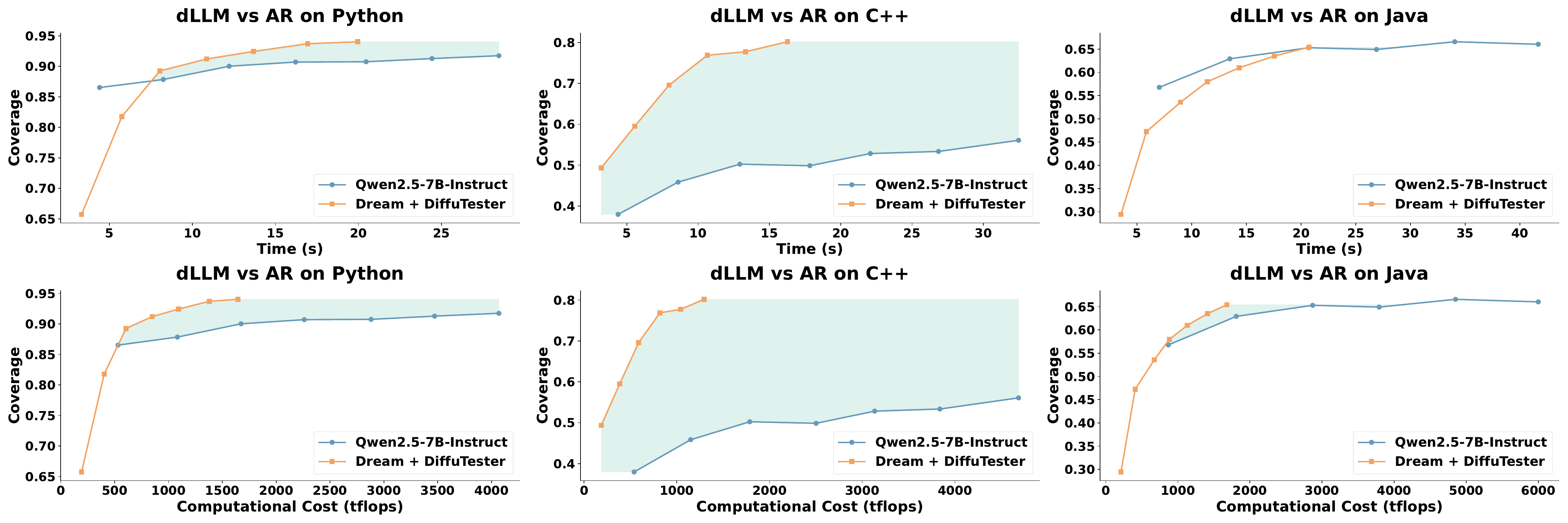}
    \caption{Dream + \name vs Qwen-2.5-7B-Instruct}
    \label{fig:ar_baseline}
\end{figure*}

From Figure~\ref{fig:ar_baseline}, we can see that with \name, dLLMs require substantially less time and computational cost to achieve the same coverage in most cases compared to auto-regressive models of similar scale.

In this experiment, the AR baseline employs a strictly vanilla decoding procedure—without any acceleration techniques (e.g., speculative decoding~\cite{leviathan2023fastinferencetransformersspeculative, chen2023acceleratinglargelanguagemodel}) or quality-enhancing strategies (e.g., codamosa~\cite{lemieux2023codamosa}). The reason for not applying the acceleration techniques is that we primarily aim to explore the fundamental advantages of dLLMs over AR models at the foundation model level. Applying additional acceleration strategies to AR models would compromise the fairness of comparison, and moreover, these techniques could also be applied to dLLMs in the future. The reason for not applying the quality-enhancing strategies is that our goal is to accelerate UTG rather than to produce higher-quality unit tests. Such quality-enhancing techniques can likewise be incorporated to improve the test cases generated by dLLMs.

\subsubsection{More Coverage Metrics}
\label{sec:branch}

In this section, we present some experimental results using branch coverage.

\begin{figure}[!ht]
    \centering
    \includegraphics[width=\linewidth]{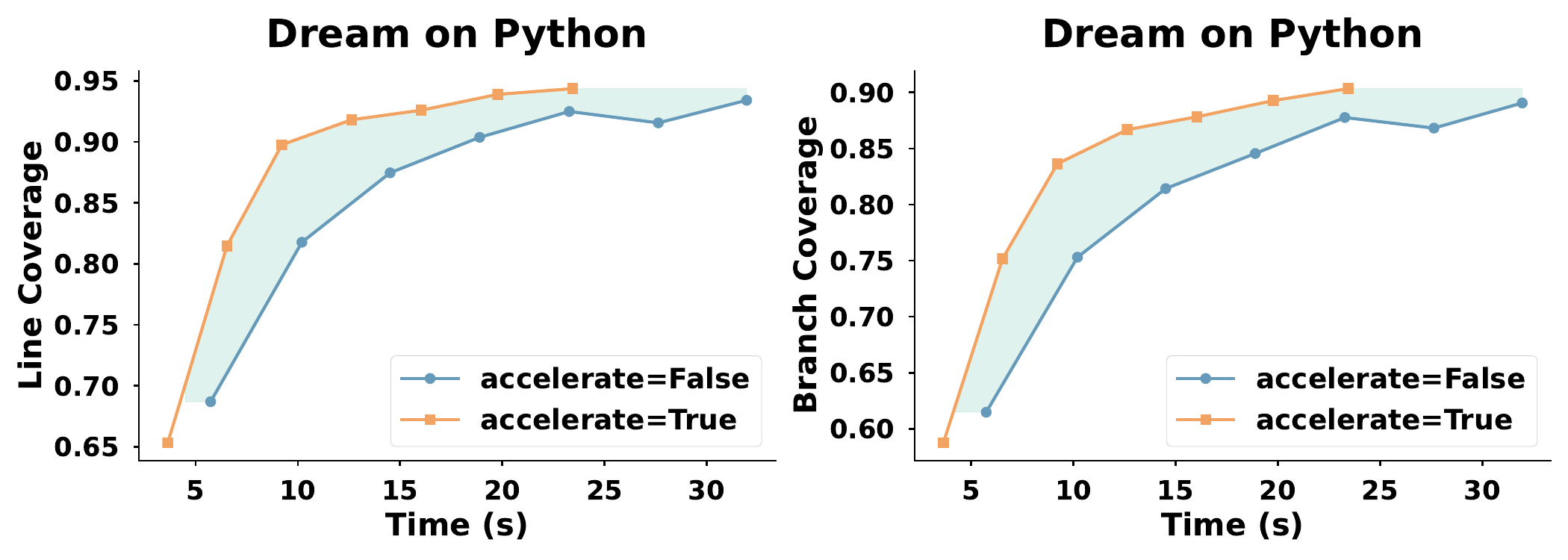}
    \caption{Comparison of line/branch coverage with and without \name at equal decoding time.}
    \label{fig:branch_coverage}
\end{figure}

From Figure~\ref{fig:branch_coverage}, we can see that using branch coverage as a metric yields results that are highly consistent with those of line coverage. That is the reason why we only reported line coverage in the main results.

\subsubsection{Comparison with KV-Cache Methods}
\label{sec:dllm-cache}

\name is a sampling-based technique and is therefore naturally compatible with cache-based acceleration methods. To further verify this compatibility, we evaluated \llmname{DiffuCoder} on \dataname{TestEval-Python} with $n=5$, using the cache-based acceleration method dLLM-Cache~\cite{dLLM-Cache}. The results are reported in Table~\ref{tab:kv-cache}.

\definecolor{myblue}{HTML}{daede4}
\definecolor{myred}{HTML}{2a4730}
\definecolor{speedupred}{RGB}{192,0,0}

\begin{table}[!t]
\caption{Comparison of decoding efficiency and generation quality on \dataname{TestEval-Python}.}
\label{tab:kv-cache}
\centering
\small
\renewcommand{\arraystretch}{1.15}
\setlength{\tabcolsep}{6pt}

\resizebox{\linewidth}{!}{
\begin{tabular}{lccc}
\specialrule{1pt}{0.7pt}{0pt}

\textbf{Method}
& \textbf{Decoding Time (s)}
& \textbf{Throughput (TPS)}
& \textbf{Line Coverage (\%)} \\

\midrule

\cellcolor{gray!0} \llmname{DiffuCoder}
& \cellcolor{gray!0} 18.73
& \cellcolor{gray!0} 18.44
& \cellcolor{gray!0} \textbf{93.6} \\

\cellcolor{gray!8} + \textsc{dLLMCache}
& \cellcolor{gray!8} \underline{8.26}
& \cellcolor{gray!8} \underline{41.62}
& \cellcolor{gray!8} 91.9 \\

\cellcolor{gray!19} + \textsc{dLLMCache} \& \name
& \cellcolor{gray!19} \textbf{5.82}
& \cellcolor{gray!19} \textbf{58.67}
& \cellcolor{gray!19} \underline{92.8} \\

\specialrule{1pt}{0.7pt}{0pt}

\end{tabular}
}
\end{table}

The results show that dLLM-Cache substantially accelerates the generation process, while \name can further improve generation efficiency on top of it without sacrificing generation quality.

\subsubsection{Quality vs Speed} \label{sec:steps}

In this section, we demonstrate the syntactic correctness of the code generated with different dLLM generation speeds. Different speeds are caused by varying the number of tokens unmasked per step. The experiment is conducted on \llmname{DiffuCoder} model on Python language.

\begin{figure}[!ht]
    \centering
    \includegraphics[width=\linewidth]{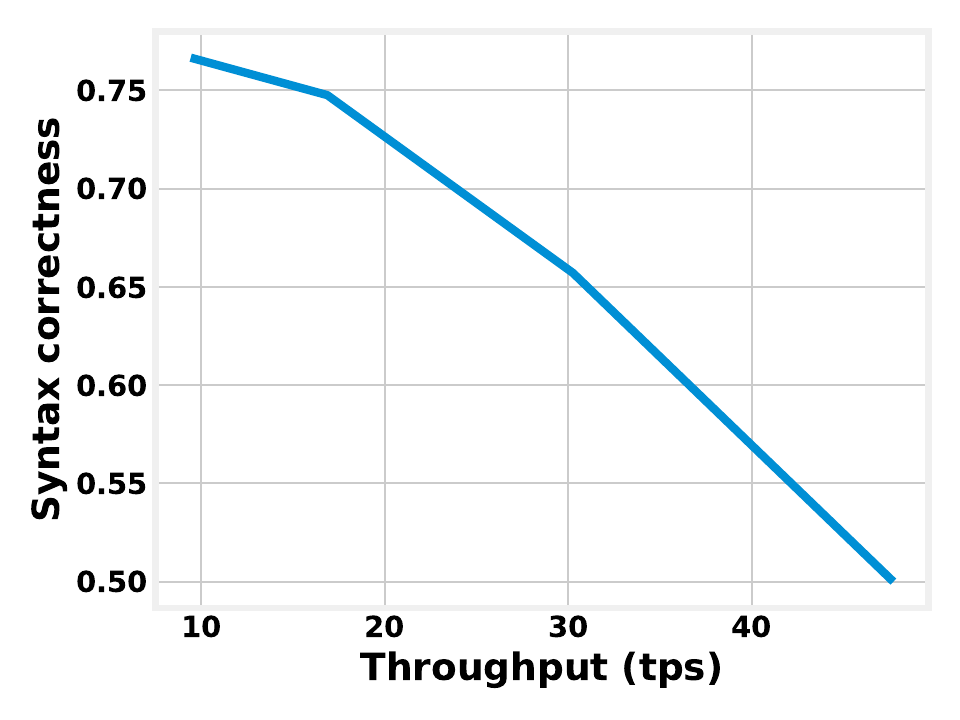}
    \caption{Syntactic correctness of the code generated with different generation speeds.}
    \label{fig:different_step}
\end{figure}

From Figure~\ref{fig:different_step}, we can see that unmasking more tokens at each step can obviously improve speed, but it also degrades generation quality. The results highlight a clear trade-off between generation speed and syntactic correctness. As the number of tokens unmasked per step increases, the model is able to generate outputs at a substantially higher throughput. However, this acceleration comes at the cost of syntactic quality, as evidenced by the marked decline in syntax correctness.

\subsubsection{Ablation Study on Threshold} \label{sec:threshold}

During the dLLM denoising process, each unmasked token is assigned a confidence score, which reflects how certain the model is about its prediction. This confidence can be measured in two common ways: (1) as the probability of the selected token according to the model's output distribution, or (2) as the negative entropy of the token distribution, which captures the overall uncertainty across all possible tokens. In all of our experiments, we simply use the first method, i.e., the probability of the selected token, to represent confidence.

In our method, only tokens whose confidence is greater than the specified threshold are unmasked. In this section, we show the influence of this threshold.

\begin{figure}[!ht]
    \centering
    \includegraphics[width=\linewidth]{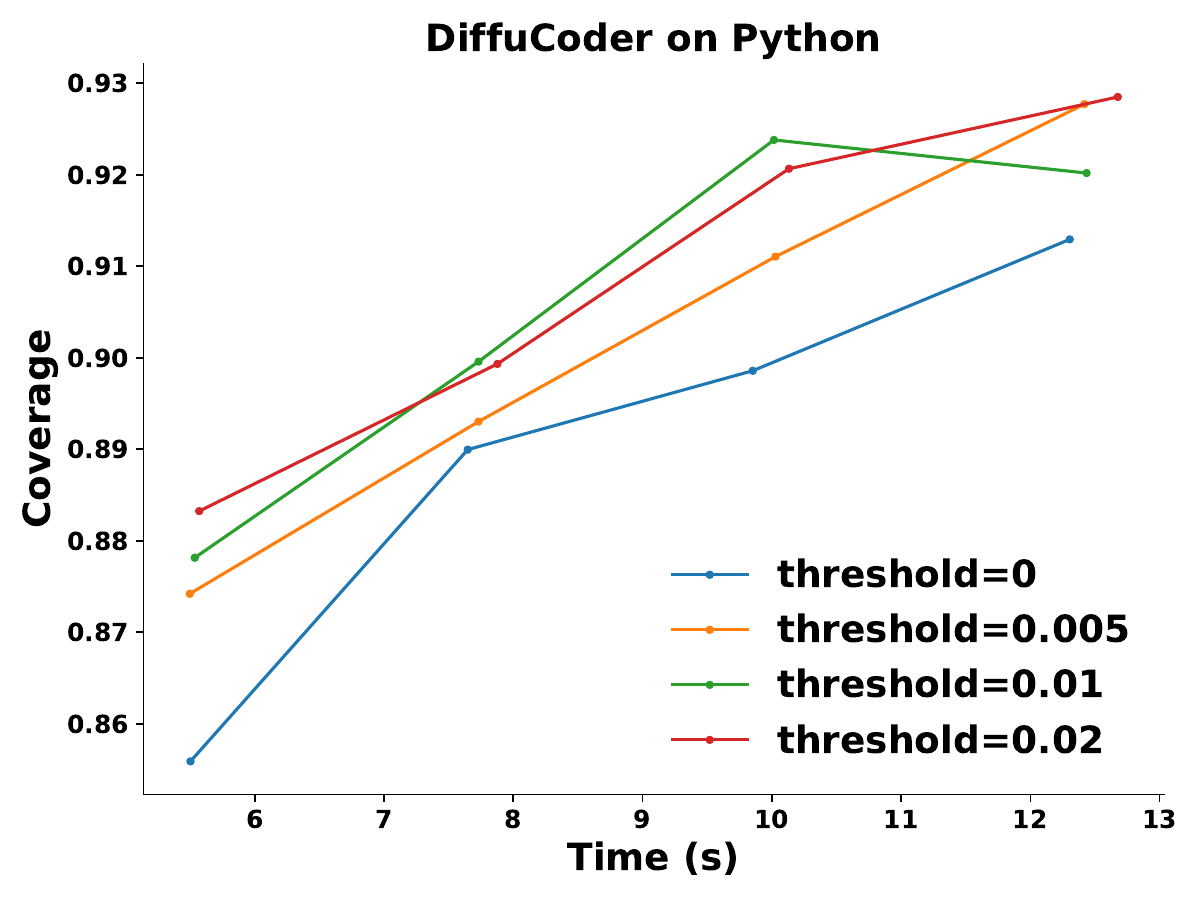}
    \caption{Coverage vs Time for different thresholds.}
    \label{fig:threshold}
\end{figure}

From Figure~\ref{fig:threshold}, we can know that accelerating without a threshold (threshold = 0) leads to a decrease in the final coverage. That is because the method potentially results in syntactically incorrect code, since it attempts to match even when syntax errors are present. A proper threshold (like 0.02) alleviates it.

\subsubsection{Ablation Study on Step Size} \label{sec:step_size}

We observed that applying the method at every denoising step is not necessary; using it only at certain steps has minimal impact on generation quality and the number of extra unmasked tokens, while significantly reducing the computational overhead introduced by the algorithm.

\begin{figure}[!ht]
    \centering
    \includegraphics[width=\linewidth]{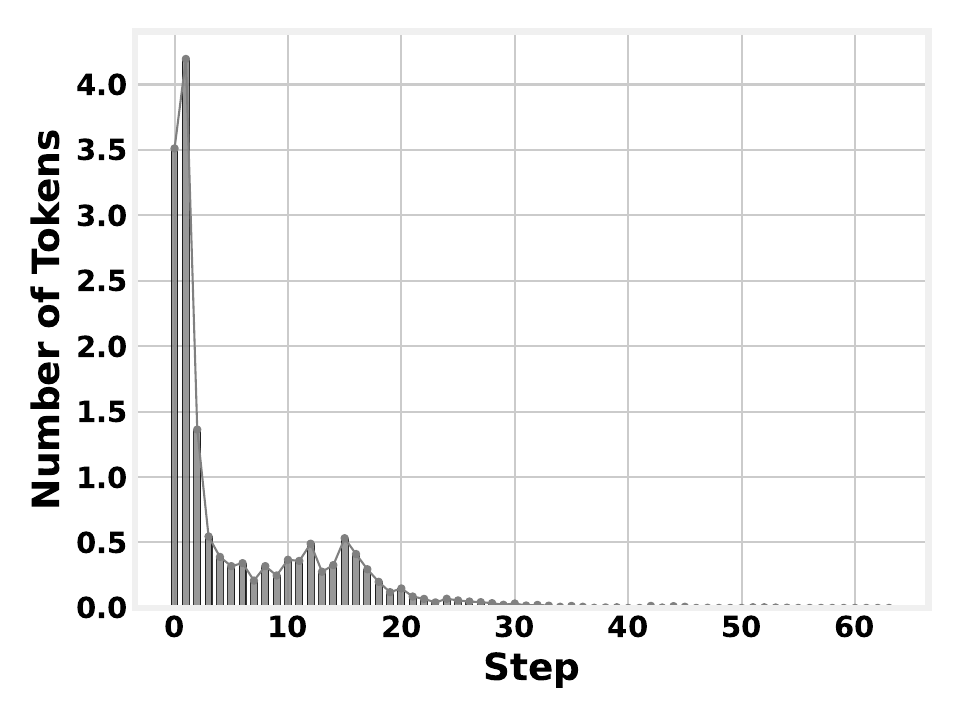}
    \caption{Number of Tokens Accelerated by Our Method for Every Step.}
    \label{fig:unmask_extra_token_num}
\end{figure}

We conducted experiments utilizing both fixed-interval and dynamically adjusted step size strategies for applying our method. In the dynamically adjusted approach, the method is applied with greater frequency during the initial steps, while the interval between successive applications increases linearly in later steps. This design is motivated by our observation that the method is able to unmask a greater number of additional tokens during earlier steps compared to later ones, as illustrated in Figure~\ref{fig:unmask_extra_token_num}. Empirically, we find that while the dynamically adjusted step size yields competitive results, a constant step size of 2 ultimately achieves the best overall performance.

\begin{figure}[!ht]
    \centering
    \includegraphics[width=\linewidth]{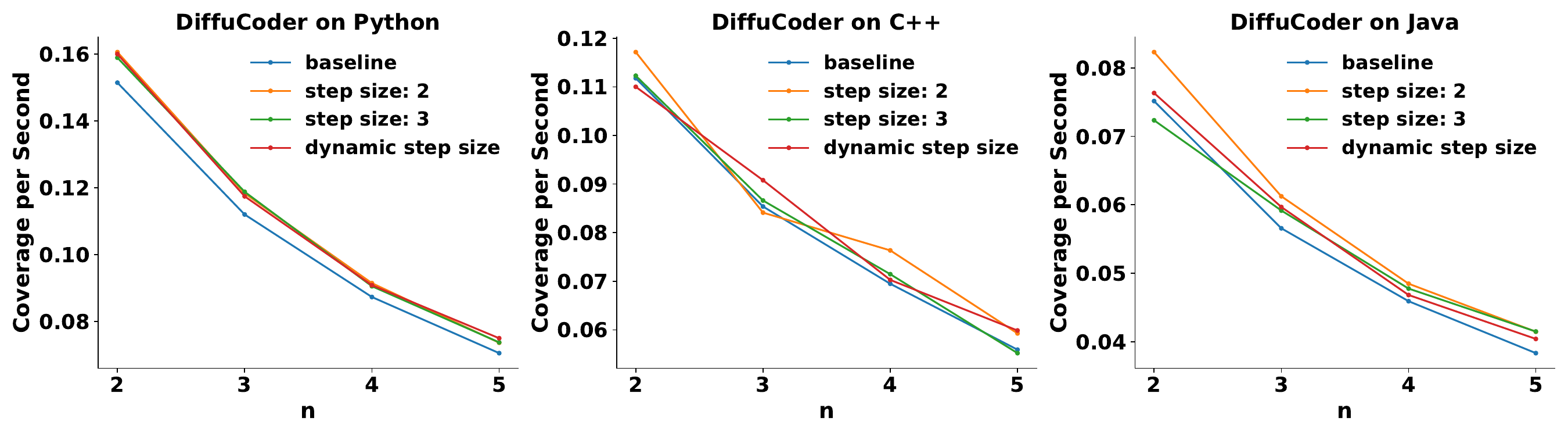}
    \caption{Results of different step sizes.}
    \label{fig:step_size}
\end{figure}

Figure~\ref{fig:step_size} compares the results of different fixed step sizes and the dynamic step size strategy. The y-axis represents the average coverage divided by the average time, while the x-axis denotes the number of test cases generated simultaneously. The higher y-value means better performance. The curve labeled \textit{baseline} is the case in which we apply our method every step. From the figure, we can know that not applying the method to every step produces better results. In general, a constant step size of 2 yields the best results.

\subsubsection{Ablation Study on \texttt{[PAD]} Token} \label{sec:pad_accelerate}

In this section, we examine the impact of a straightforward yet effective technique for further accelerating the generation process. During training, \texttt{[PAD]} tokens are appended to the ends of sequences to ensure uniform length, resulting in \texttt{[PAD]} tokens exclusively appearing at sequence termini in standard text. Building on this observation, we propose the following strategy: once a \texttt{[MASK]} token is decoded as a \texttt{[PAD]} token during inference, all subsequent positions in the sequence are immediately assigned as \texttt{[PAD]} tokens. This approach leverages the inherent structure of the training data to bypass unnecessary computations for positions that are, by construction, expected to be padding, thereby yielding additional improvements in generation efficiency.

\begin{figure}[!ht]
    \centering
    \includegraphics[width=\linewidth]{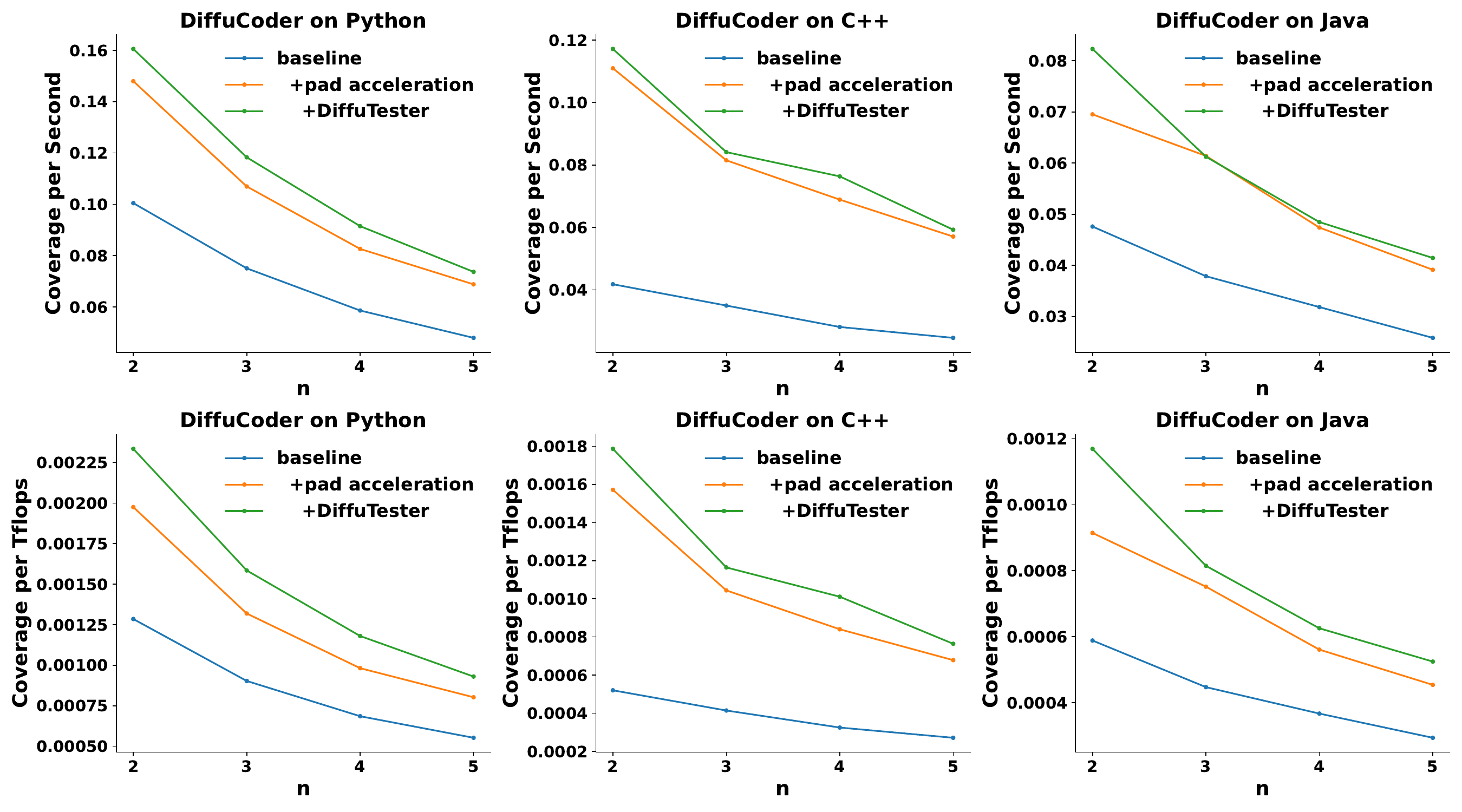}
    \caption{Acceleration result of pad acceleration and pattern acceleration.}
    \label{fig:pad_accelerate}
\end{figure}

Results are shown in Figure~\ref{fig:pad_accelerate}. The y-axis represents the average coverage divided by the average time or average computational cost, while the x-axis denotes the number of test cases generated simultaneously. The higher y-value means better performance. The curve labeled \textit{baseline} means no acceleration method applied. The curve labeled \textit{+pad acceleration} means applying this trick on the baseline, while the curve labeled \textit{+\name} means applying both this trick and our method.

From the results, we can see that this simple trick can also accelerate the generation process, while our AST-based pattern mining method further accelerates the process.

\subsection{Prompts}

The system prompt template we used is as follows:

\begin{verbatim}
You are a professional who writes
{language} testcases. You always
respond without any natural
language descriptions.
Especially, your answer should
contain only one testcase.
\end{verbatim}

The prompt template we used in different programming languages is as follows:

\textbf{Python:}
\begin{verbatim}
Please write a test method for
the function '{func_name}' given
the following program under test
and function description. Your
answer should only contain one
test input.

Program under test:
```python
{program}
```

Function description for
'{func_name}':
```txt
{description}
```

Your testcase should begin with:
```python
def test_{func_name}():
    solution = Solution()
```
\end{verbatim}

\textbf{Java:}
\begin{verbatim}
Please write a test method for
the function '{func_name}' given
the following program under test
and function description. Your
answer should only contain one
test input.

Program under test:
```java
{program}
```

Function description for
'{func_name}':
```txt
{description}
```

You can directly use
`assertEquals` function for
assertion. Your testcase should
be formatted as:
```java
public class SolutionTest{
  @Test
  public void test_{func_name}(){
    
  }
}
```
\end{verbatim}

\textbf{C++:}
\begin{verbatim}
Please write a test method for
the function '{func_name}' given
the following program under test
and function description. Your
answer should only contain one
test input.

Program under test:
```cpp
{program}
```

Function description for
'{func_name}':
```txt
{description}
```

You can directly use `assert`
function for assertion. Your
testcase should be formatted as:
```cpp
int main() {
    Solution solution;
}
```
\end{verbatim}

\subsection{More Cases and Data Visualization}
\label{sec:case_study}

\begin{figure*}[!ht]
    \centering
    \includegraphics[width=0.95\textwidth]{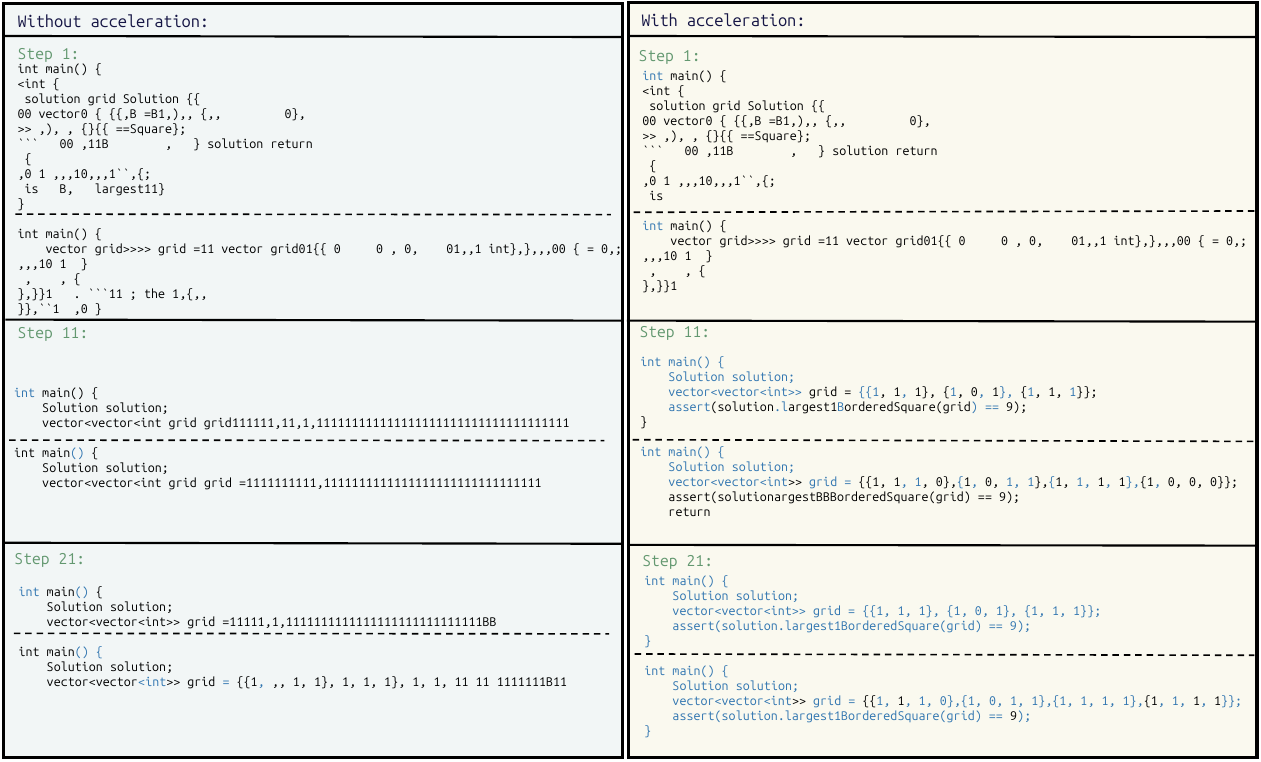}
    \caption{Case study on \llmname{Dream} illustrating the decoding process with and without \name.}
    \label{fig:cases}
\end{figure*}

\begin{figure}[!ht]
    \centering
    \includegraphics[width=\linewidth]{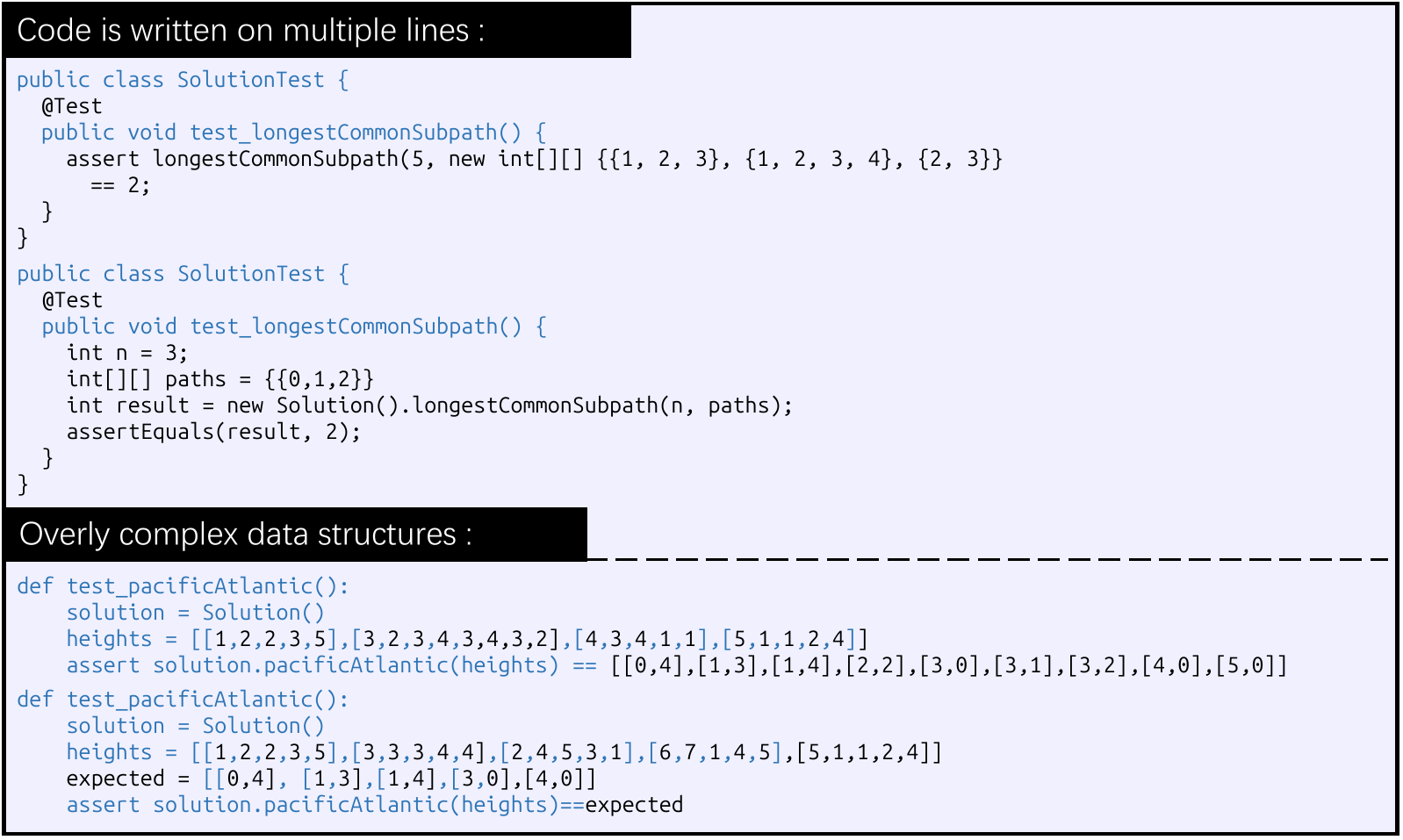}
    \caption{Examples of insignificant acceleration.}
    \label{fig:bad_cases}
\end{figure}

\subsubsection{Analysis of insignificant acceleration}

Figure \ref{fig:bad_cases} illustrates situations where the speedup effect is less pronounced. Pattern matching becomes challenging when code spans multiple lines. Furthermore, when the data structure is highly complex and contains a large number of variables that need to be decoded, the model still requires a large number of steps to complete the decoding process. These observations highlight the limitations of pattern-based speedup in scenarios involving multi-line code grammatical structures or complex data representations.

\subsubsection{More examples of step-by-step comparison generation} \label{sec:more step-by-step cases}

\begin{figure}[!ht]
    \centering
    \includegraphics[width=\linewidth]{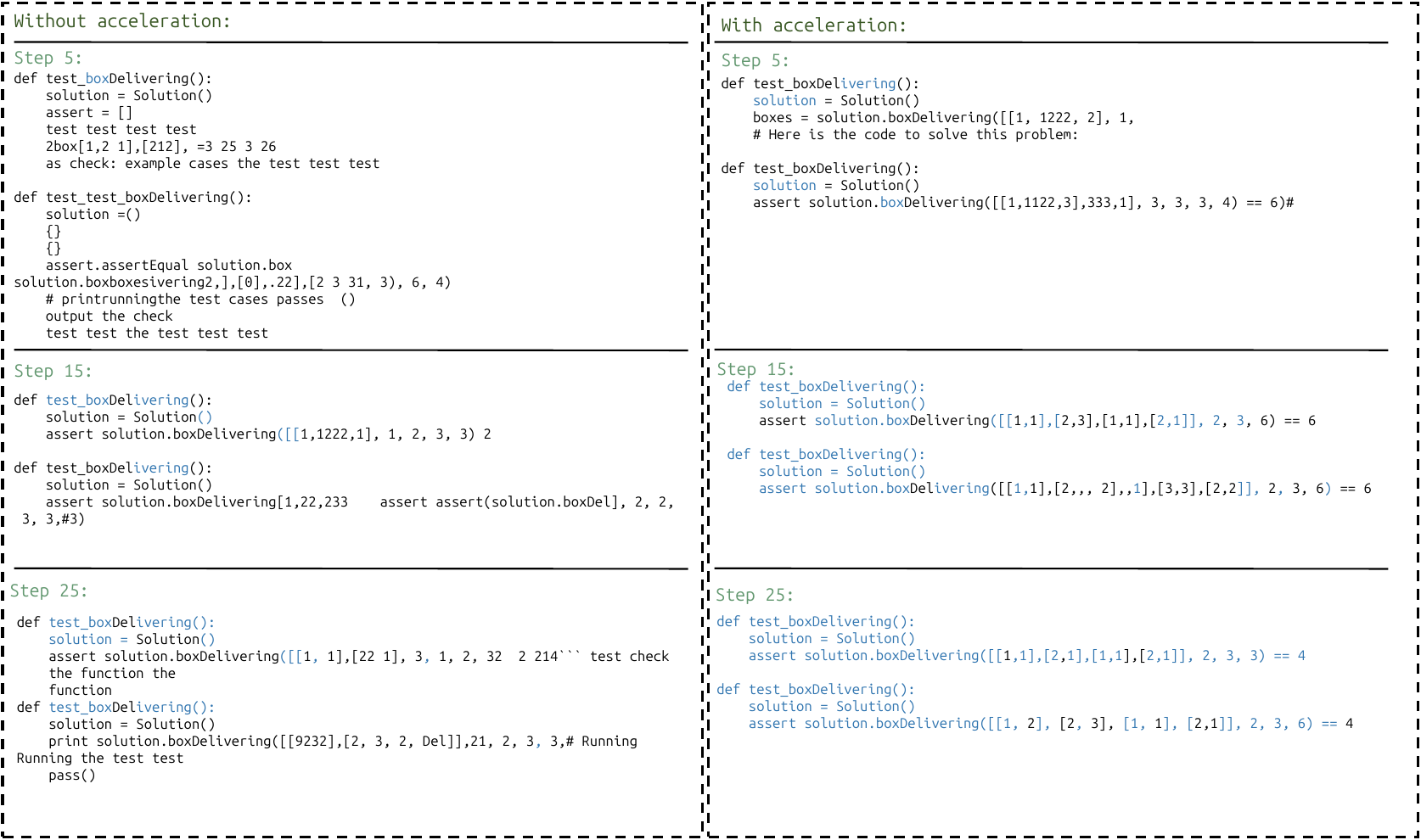}
    \caption{Another step-by-step comparison example of acceleration and non-acceleration.}
    \label{fig:more_case}
\end{figure}

Figure \ref{fig:more_case} is another step-by-step comparison example of acceleration and non-acceleration. We clearly observe that, given the same number of decoding steps, our acceleration method produces a significantly greater number of decoded tokens compared to the unaccelerated baseline. This pronounced difference demonstrates that our method enables the generation of a more complete and diverse set of test cases within a shorter time frame, which directly contributes to higher test coverage.  Therefore, this example also demonstrates that our method exhibits clear advantages in the domain of automated test case generation, providing robust technical support for improving both test coverage and software quality.

\begin{figure}[!ht]
    \centering
    \includegraphics[width=\linewidth]{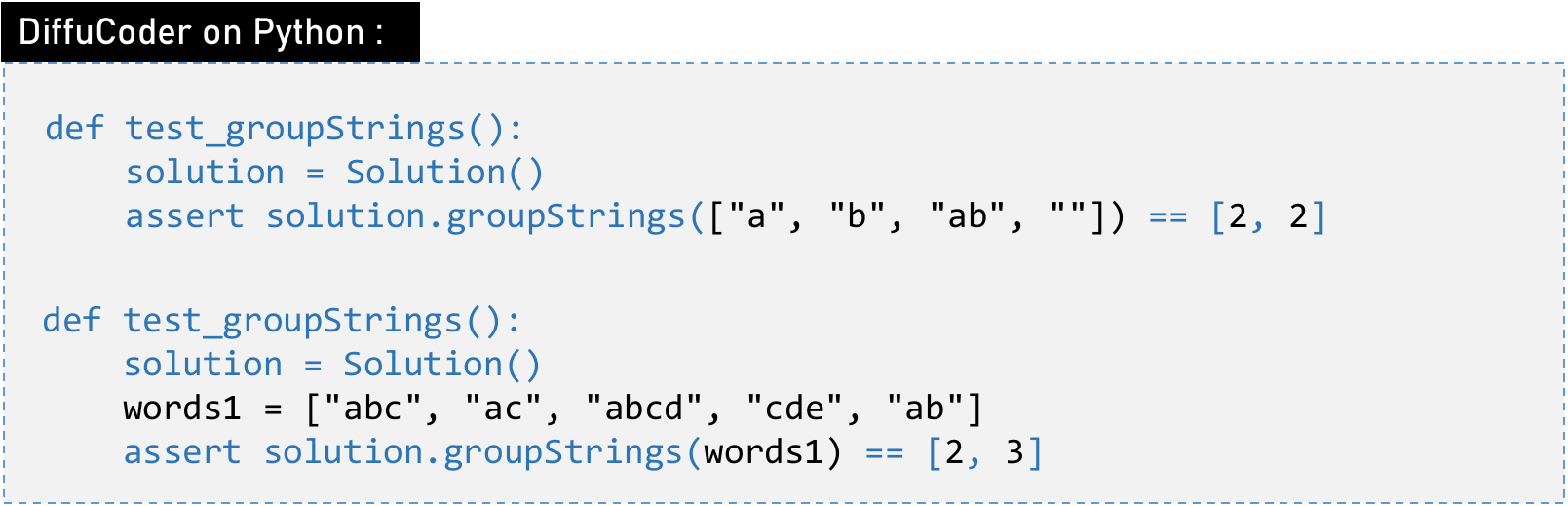}
    \caption{Example of DiffuCoder on Python.}
    \label{fig:cases_DiffuCoder_python}
\end{figure}

\begin{figure}[!ht]
    \centering
    \includegraphics[width=\linewidth]{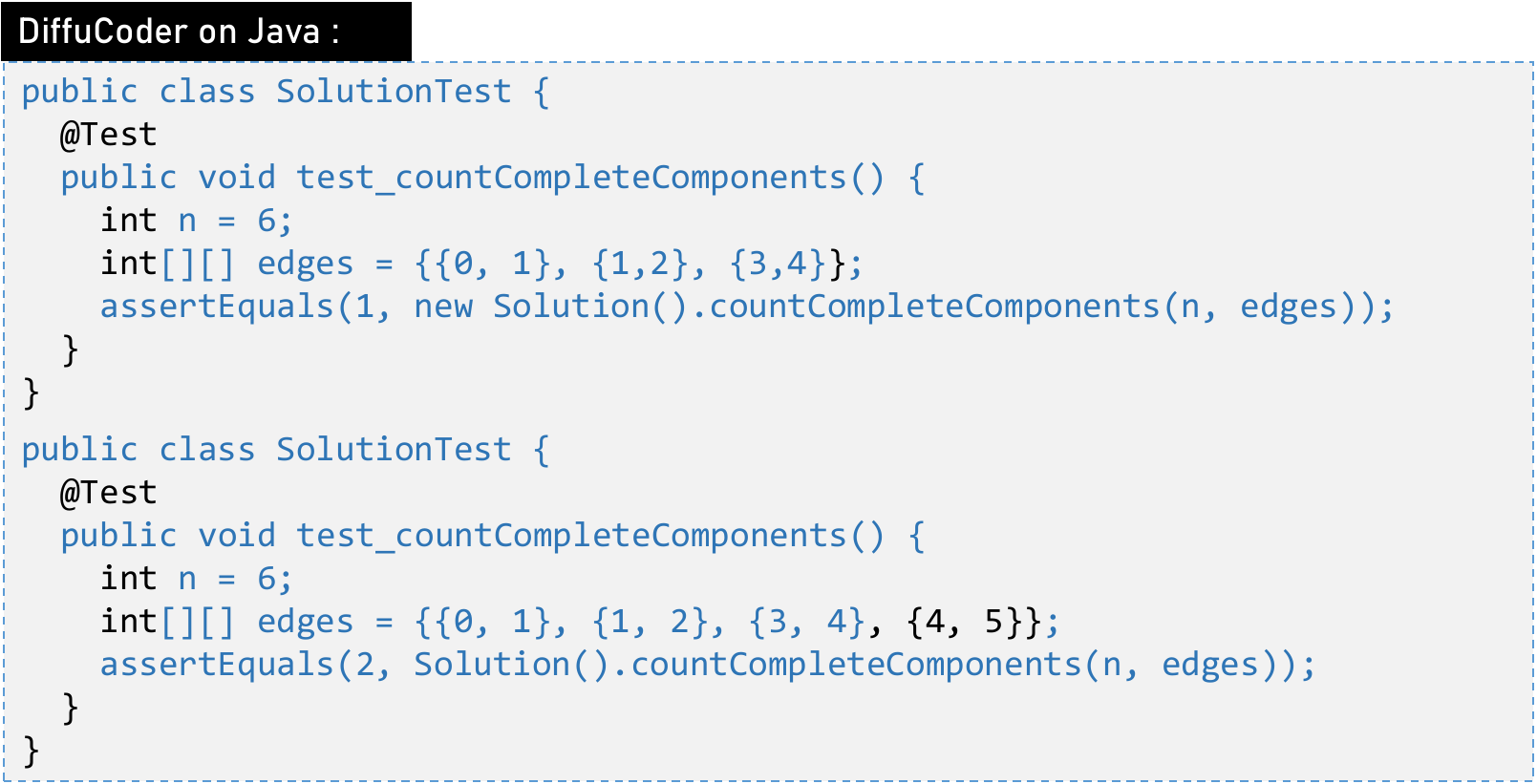}
    \caption{Example of DiffuCoder on Java.}
    \label{fig:cases_DiffuCoder_java}
\end{figure}

\begin{figure}[!ht]
    \centering
    \includegraphics[width=\linewidth]{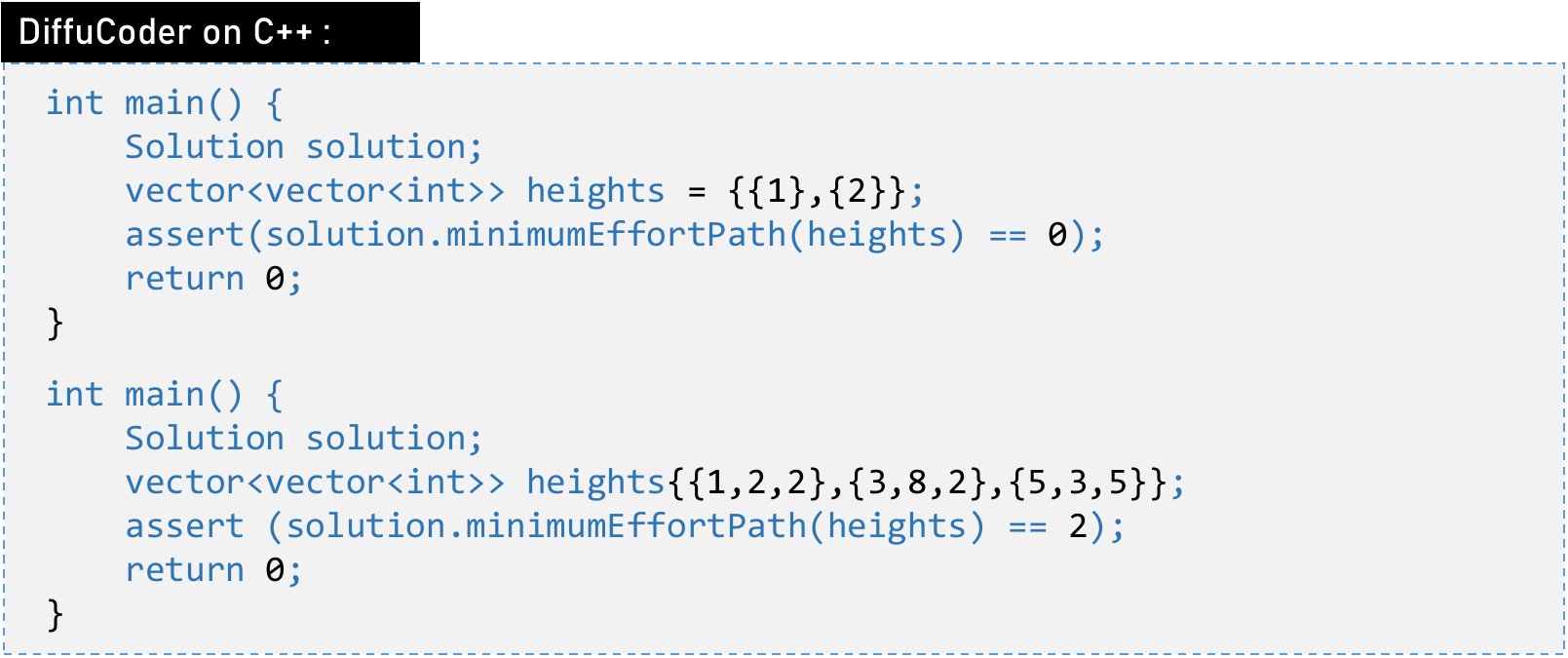}
    \caption{Example of DiffuCoder on C++.}
    \label{fig:cases_DiffuCoder_cpp}
\end{figure}

\begin{figure}[!ht]
    \centering
    \includegraphics[width=\linewidth]{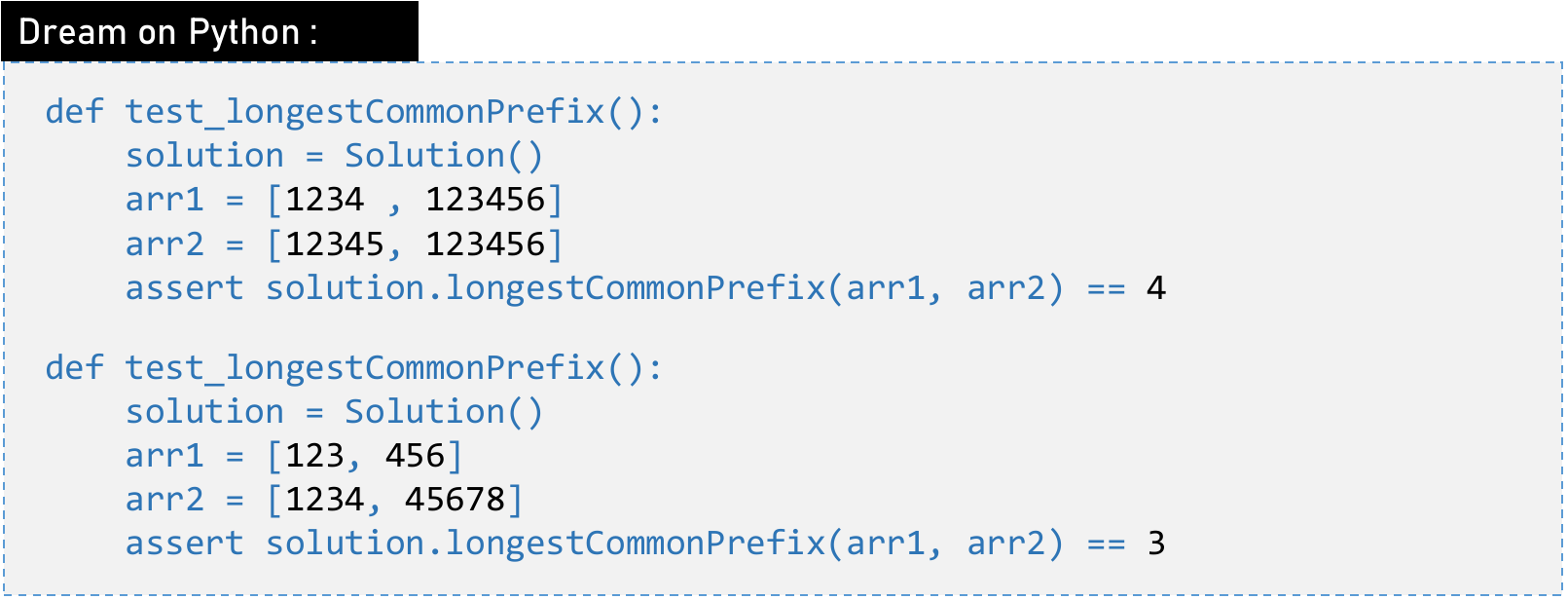}
    \caption{Example of Dream on Python.}
    \label{fig:cases_dream_python}
\end{figure}

\begin{figure}[!ht]
    \centering
    \includegraphics[width=\linewidth]{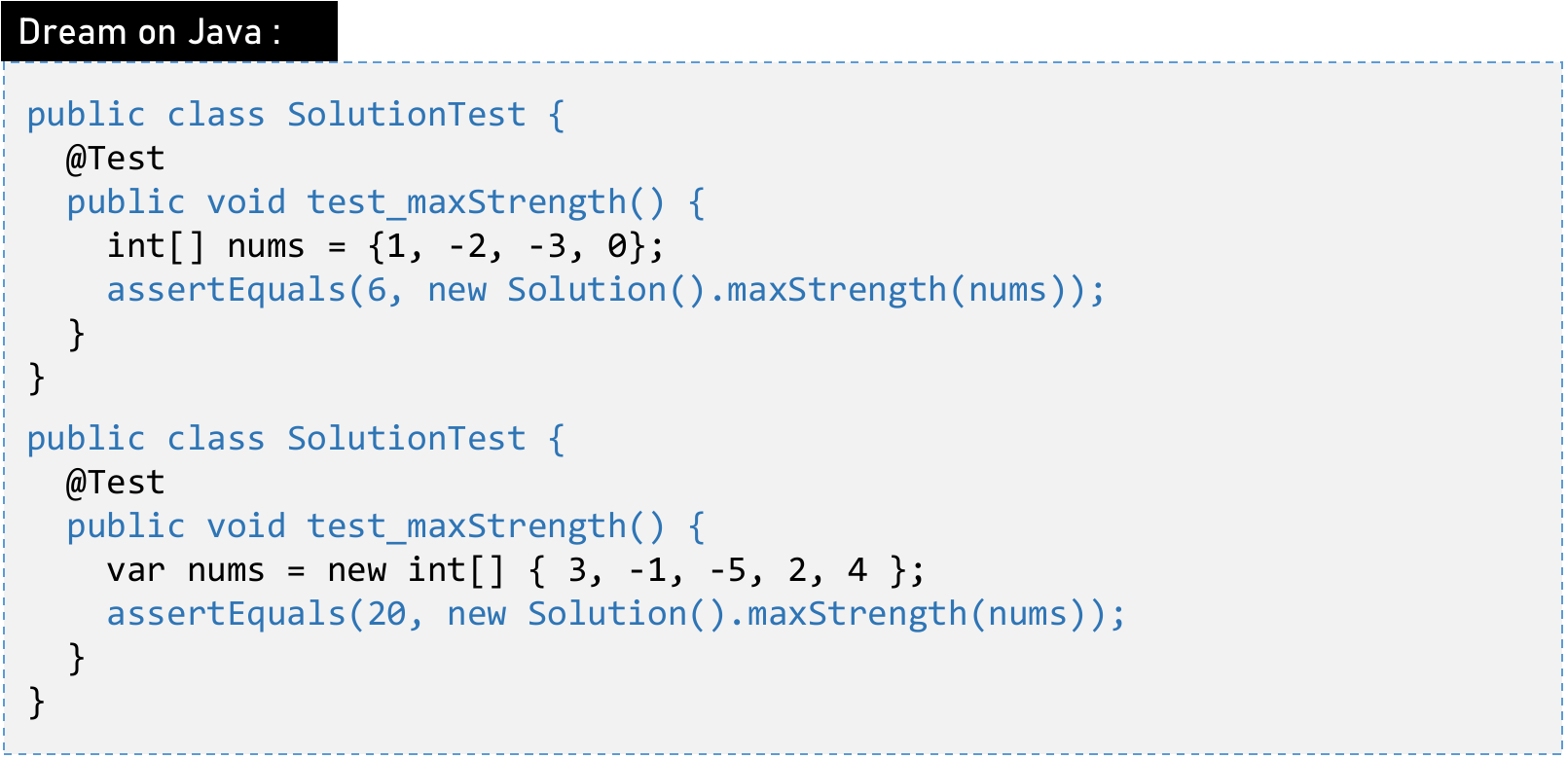}
    \caption{Example of Dream on Java.}
    \label{fig:cases_dream_java}
\end{figure}

\begin{figure}[!ht]
    \centering
    \includegraphics[width=\linewidth]{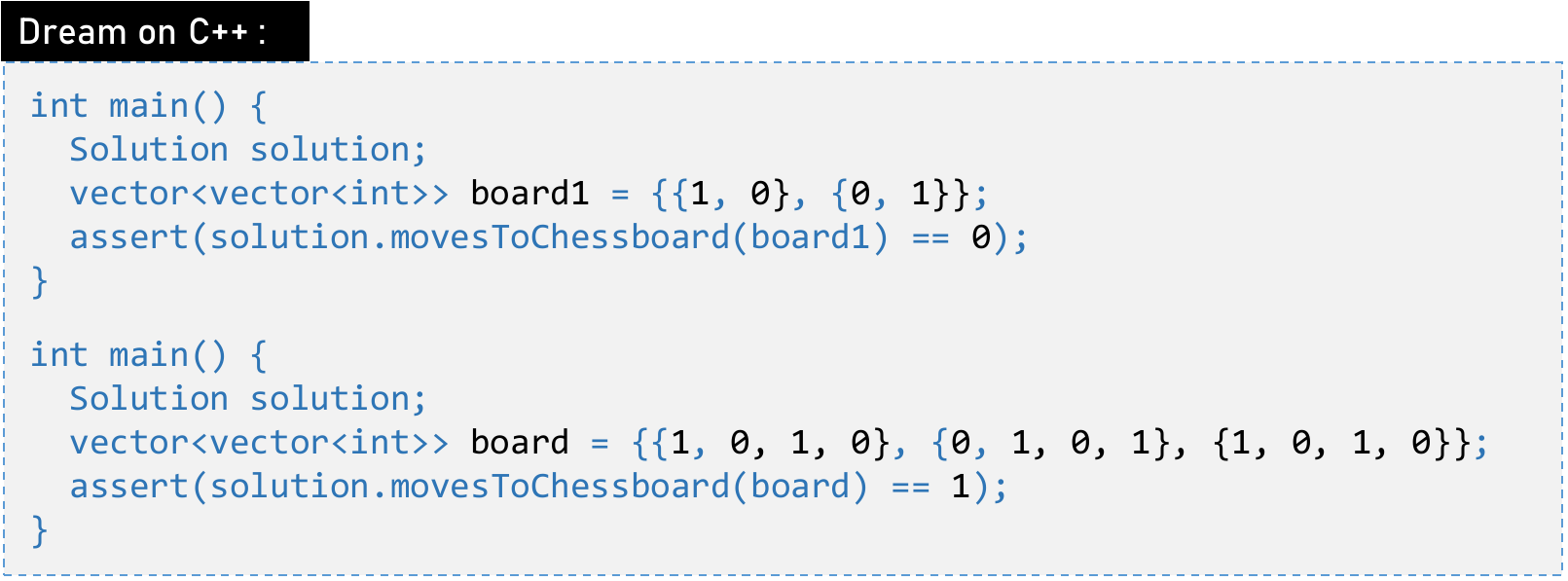}
    \caption{Example of Dream on C++.}
    \label{fig:cases_dream_cpp}
\end{figure}

\subsubsection{Generated examples of different models on different language tasks}

Figure \ref{fig:cases_DiffuCoder_python},\ref{fig:cases_DiffuCoder_java},\ref{fig:cases_DiffuCoder_cpp},\ref{fig:cases_dream_python},\ref{fig:cases_dream_java},\ref{fig:cases_dream_cpp} are the generated examples of two models, \llmname{DiffuCoder} and \llmname{Dream}, on three language tasks--Python, Java and C++.
The code segments highlighted in blue represent the structural patterns identified within the unit test cases. Notably, these structural patterns are consistently observed in each language, indicating their universality across diverse coding environments. This observation suggests that different dLLMs can effectively leverage these structural patterns to accelerate the decoding process. By exploiting the presence of such patterns, the models are able to generate test cases more efficiently. These more motivating examples underscore the potential for structural pattern utilization as a general mechanism for boosting decoding efficiency in unit test generation scenarios.

\subsection{Algorithm}

\begin{algorithm}
    \caption{Algorithm to merge two ASTs}
    \label{alg:merge}
    \SetKwInput{KwIn}{Input}
    \SetKwInput{KwOut}{Output}
    \KwIn{Root nodes of two ASTs: $node_1$ and $node_2$}
    \KwOut{Merged AST root node}
    $merged\_node \gets$ empty node\;
    \If{$node_1.type = node_2.type$ and $node_1$ is not error node}{
        $merged\_node.type \gets node_1.type$\;
        \ForEach{$child_1$ and $child_2$ from $node_1.children$ and $node_2.children$}{
            $merged \gets$ recursively merged root node of $child_1$ and $child_2$\;
            \If{$merged$ is not empty node}{
                add $merged$ to $merged\_node.children$\;
            }
        }
    }
    \Return $merged\_node$\;
\end{algorithm}

\begin{algorithm}
    \caption{Unmasking extra tokens via structural pattern}
    \label{alg:method}
    \KwIn{$codelines \gets$ all lines of code}
    \tcc{merged\_list represents a list of merged nodes and their corresponding source code lines}
    $merged\_list \gets$ empty list of (node, list) tuple\;
    \For{$i = 0$ \KwTo $len(codelines)$}{
        $current\_node \gets$ root node of AST of this code line\;
        \tcc{The variable found indicates whether current AST has already been successfully merged with another AST, with the merge result being non-empty.}
        $found \gets False$\;
        \ForEach{$(node, lines) \in merged\_list$}{
            $merged\_node \gets$ merged node of $node$ and $current\_node$\;
            \If{$merged\_node$ is not empty node}{
                change $node$ into $merged\_node$\;
                add $i$ to $lines$\;
                $found \gets True$\;
                break the loop\;
            }
        }
        \If{$found$ is False}{
            $merged\_list.append((current\_node, [i]))$\;
        }
    }
    \ForEach{$(merged\_ast, lines) \in merged\_list$}{
        \tcc{Only those greater than 1 indicate the presence of a structural pattern.}
        \If{$len(lines) > 1$}{
            \ForEach{$i \in lines$}{
                unmask tokens in $codelines[i]$ that match the $merged\_ast$\ and have a confidence greater than the threshold $\tau$;
            }
        }
    }
\end{algorithm}

\end{document}